\documentclass[twocolumn,appendixfloats,tighten]{aastex631}
\usepackage[T1]{fontenc}
\usepackage{ae,aecompl}

\usepackage{amsmath}	
\usepackage{amssymb}	

\usepackage{ctable}
\usepackage{url}
\usepackage{xspace}
\usepackage[normalem]{ulem}
\usepackage{hyperref}
\hypersetup{linkcolor=red,citecolor=blue,filecolor=cyan,urlcolor=magenta}
\usepackage[all]{hypcap}
\usepackage{graphicx}
\usepackage{comment}
\usepackage{color}

\usepackage{amssymb}
\usepackage{pifont}

\usepackage{etoolbox}
\newtoggle{ApJFigs}
\togglefalse{ApJFigs}

\def\msun{{\rm\,M_\odot}}

\newcommand{\kms}{\, {\rm km\, s}^{-1}}

\newcommand{\be}{\begin{equation}}
\newcommand{\ee}{\end{equation}}

\newcommand{\au}{\mathrm{au}}

\newcommand{\F}{Fig.}

\newcommand{\eq}{equation}

\newcommand{\myr}{\mathrm{Myr}}
\newcommand{\yr}{\mathrm{yr}}
\newcommand{\gyr}{\mathrm{Gyr}}

\newcommand{\mse}{\textsc{MSE}}

\begin{document}

\title{A statistical view on stable and unstable Roche lobe overflow of a tertiary star onto the inner binary in triple systems} 

\shortauthors{Hamers et al.}

\author[0000-0003-1004-5635]{Adrian S. Hamers}
\affiliation{Max-Planck-Institut f\"{u}r Astrophysik, Karl-Schwarzschild-Str. 1, 85741 Garching, Germany}

\author[0000-0002-6012-2136]{Hila Glanz}
\affiliation{Technion - Israel Institute of Technology, Haifa, 3200002, Israel}

\author[0000-0001-5853-6017]{Patrick Neunteufel}
\affiliation{Max-Planck-Institut f\"{u}r Astrophysik, Karl-Schwarzschild-Str. 1, 85741 Garching, Germany}

\begin{abstract} 
In compact stellar triple systems, an evolved tertiary star can overflow its Roche lobe around the inner binary. Subsequently, the tertiary star can transfer mass to the inner binary in a stable manner, or Roche lobe overflow (RLOF) can be unstable and lead to common-envelope (CE) evolution. In the latter case, the inner binary enters the extended envelope of the tertiary star and spirals in towards the donor's core, potentially leading to mergers or ejections. Although studied in detail for individual systems, a comprehensive statistical view on the various outcomes of triple RLOF is lacking. Here, we carry out $10^5$ population synthesis simulations of tight triples, self-consistently taking into account stellar evolution, binary interactions, and gravitational dynamics. Also included are prescriptions for the long-term evolution of stable triple mass transfer, and triple CE evolution. Although simple and ignoring hydrodynamic effects, these prescriptions allow for a qualitative statistical study. We find that triple RLOF occurs in $\sim 0.06\%$ of all triple systems. Of these 0.06\%, $\sim 64\%$ of cases lead to stable mass transfer, and $\sim 36\%$ to triple CE evolution. Triple CE is most often ($\sim 76\%$) followed by one or multiple mergers in short succession, most likely an inner binary merger of two main-sequence stars. Other outcomes of triple CE are a binary+single system ($\sim 23\%$, most of which not involving exchange interactions), and a stable triple ($\sim 1\%$). We also estimate the rate of Type Ia supernovae involving white dwarf mergers following triple RLOF, but find only a negligible contribution. 
\end{abstract}

\section{Introduction}
\label{sect:intro}
Triple systems in hierarchical configurations are well known for Zeipel-Lidov-Kozai (ZLK) oscillations, which arise in these systems from long-term gravitational exchanges of angular momentum between the inner and outer orbits (\citealt{1910AN....183..345V,1962P&SS....9..719L,1962AJ.....67..591K}; see \citealt{2016ARA&A..54..441N,2017ASSL..441.....S,2019MEEP....7....1I} for reviews). ZLK oscillations, or, more generally, secular evolution, can periodically produce high eccentricities in the inner orbit, with important implications depending on the astrophysical setting. These implications include (in combination with tidal effects) the production of short-period binaries in stellar triples (e.g., \citealt{1979A&A....77..145M,1998MNRAS.300..292K,2001ApJ...562.1012E,2006Ap&SS.304...75E,2007ApJ...669.1298F,2014ApJ...793..137N}), and the production of hot Jupiters in binary-star systems (e.g., \citealt{2003ApJ...589..605W,2007ApJ...669.1298F,2012ApJ...754L..36N,2015ApJ...799...27P,2016MNRAS.456.3671A,2016ApJ...829..132P,2017ApJ...835L..24H}). Other implications are enhancing mergers of compact objects leading to gravitational wave (GW) events (e.g., \citealt{2002ApJ...578..775B,2011ApJ...741...82T,2013MNRAS.430.2262H,2014MNRAS.439.1079A,2017ApJ...841...77A,2017ApJ...836...39S,2017ApJ...846L..11L,2018ApJ...863...68L,2018ApJ...865....2H,2018ApJ...856..140H,2018ApJ...853...93R,2018ApJ...864..134R,2018A&A...610A..22T,2019MNRAS.486.4443F}), affecting the evolution of protoplanetary or accretion disks in binaries (e.g., \citealt{2014ApJ...792L..33M,2015ApJ...813..105F,2017MNRAS.467.1957Z,2017MNRAS.469.4292L,2018MNRAS.477.5207Z,2019MNRAS.485..315F,2019MNRAS.489.1797M}), triggering white dwarf (WD) pollution by planets (e.g., \citealt{2016MNRAS.462L..84H,2017ApJ...834..116P}), and producing blue straggler stars (e.g., \citealt{2009ApJ...697.1048P,2016ApJ...816...65A,2016MNRAS.460.3494S,2019MNRAS.488..728F}).

The implications of strong secular evolution (i.e., eccentricity excitation) in stellar triples have been given much attention in the past decades, with also the effects of stellar and tidal evolution taken into account \citep[e.g.,][]{2013MNRAS.430.2262H,2016ComAC...3....6T,2016MNRAS.460.3494S,2019ApJ...878...58S}. However, stellar triple systems can give rise to other types of interesting evolution when not only secular and stellar evolution are included, but also other effects such binary interactions including common-envelope (CE) evolution, short-term dynamical instabilities, and other processes. Such evolution can produce unusual systems and transient phenomena that are otherwise hard to explain with isolated binary evolution. For example, mass changes due to stellar evolution or binary interactions can destabilize triples, and such instabilities can lead to highly eccentric mergers between stars with different evolutionary stages, as well as ejections of stars \citep{1994MNRAS.270..936K,1999ApJ...511..324I,2010arXiv1001.0581P,2011MNRAS.412.2763F,2011ApJ...734...55P,2012ApJ...760...99P,2021arXiv210713620H,2021arXiv210804272T}. This includes producing eccentric post-CE binaries with evolved or WD components such as the Sirius WD-main-sequence (MS) binary \citep[e.g.,][]{2005ApJ...630L..69L,2008A&A...480..797B,2012ApJ...760...99P,2017ApJ...840...70B,2021MNRAS.507.2659G}. 

Another type of interaction in evolving triple stars that has received relatively little attention is `triple Roche lobe overflow' (RLOF), during which the tertiary star overflows its Roche lobe around the more compact inner binary \citep[e.g.,][]{1980IAUS...88..145M,1996ASIC..477..345E,2014ApJ...781L..13T,2020MNRAS.491..495D,2020MNRAS.493.1855D}. This type of evolution is expected to occur in a subset of triple systems in which the tertiary star is the initially most massive star, and the outer orbit is sufficiently compact in order to fill its Roche lobe around the inner binary. In $\sim 18\%$ of triples in the Solar neighborhood, the tertiary star is indeed the most massive star in the system \citep{2010yCat..73890925T,2014MNRAS.438.1909D}. 

Triple RLOF can lead to stable mass transfer onto the inner binary system during which some of the mass may be accreted by the inner binary components and some is ejected from the system \citep[e.g.,][]{2014MNRAS.438.1909D,2019ApJ...871...84M,2020ApJ...889..114M}, or to unstable transfer which is expected to lead to CE-like behavior, with the inner binary being engulfed by the tertiary star's extended envelope \citep[e.g.,][]{2015MNRAS.450.1716S,2020MNRAS.498.2957C,2021MNRAS.500.1921G,2021MNRAS.505.4791S}. In the unstable case, the inner binary spirals in towards the donor's core due to friction; if migration is efficient enough, then the inner binary becomes dynamically unstable with respect to the donor's core, triggering potentially strong and chaotic three-body interactions during which collisions could occur, or components could be ejected from the system. 

Another type of CE evolution in triple systems is the `circumbinary' case, when the inner binary contains the most massive star and first initiates CE evolution with its inner binary component, but the donor star's envelope swells up sufficiently to also engulf the tertiary star \citep{2021MNRAS.500.1921G}. 

There exists a large variety of possibilities involving the generally complicated process of triple RLOF (leading to either stable or unstable mass transfer), and most attention has been given to detailed simulations of individual systems \citep[e.g.,][]{2014MNRAS.438.1909D,2017MNRAS.471.3456H,2019MNRAS.490.4748S,2021MNRAS.500.1921G}. Few studies have focussed on statistical aspects of triple RLOF for a population of stellar triples, especially with self-consistently taking into account important processes such as stellar and binary evolution, and dynamics. These aspects include the frequency of triple RLOF, and its outcomes.

Some statistical aspects of triple RLOF were considered by \citet{2020MNRAS.496.1819L}, who focused on the production of nearly equal-mass binaries through stable triple mass transfer (motivated by the detailed simulations of \citealt{2019ApJ...876L..33P}). \citet{2020MNRAS.496.1819L} focused on specific aspects of stable triple RLOF, and, by using a binary population synthesis code, did not take into account any coupling of the evolution between the inner binary and tertiary star until the onset of triple RLOF. Here, we carry out a more self-consistent study in which we use the population synthesis code \mse~\citep{2021MNRAS.502.4479H}, which allows us to simultaneously take into account stellar evolution, binary interactions such as tidal evolution, mass transfer and CE evolution (in both inner and outer orbits), and gravitational dynamics (including secular evolution and hence ZLK eccentricity excitation). We also take into account both phases of stable triple mass transfer as well as triple CE evolution (the `circumstellar' case), by adopting simple prescriptions for these phases. 

This paper is structured as follows. We briefly describe the population synthesis code used in this work in \S\ref{sect:meth}, focussing mainly on the assumed prescriptions for modelling triple stable mass transfer and triple CE evolution. We give a number of examples of different types of evolution involving triple RLOF in \S\ref{sect:ex} (these examples are based on population synthesis calculations). In \S\ref{sect:ICs}, we describe the assumptions made to generate a population of triples that we evolve with \mse~in order to study statistical aspects of triple RLOF; the latter are presented in \S\ref{sect:popsyn}. We give a discussion in \S\ref{sect:dis}, and conclude in \S\ref{sect:conclusions}.

\section{Methodology}
\label{sect:meth}
Here, we briefly describe the main ingredients of the population synthesis code used in this work, \mse\footnote{The version used for the population synthesis simulations in this work is v0.84.}. This code can handle a system of any multiplicity (and including planetary subsystems), although it is here applied exclusively to stellar triples. For more details on \mse, we refer to \citet{2021MNRAS.502.4479H}.

\subsection{Dynamics}
\label{sect:meth:dyn}
Gravitational dynamical evolution of the system is modelled via two methods: (1) secular (orbit-averaged) integration for sufficiently hierarchical systems, based on the formalism of \citet{2016MNRAS.459.2827H,2018MNRAS.476.4139H,2020MNRAS.494.5492H}, and (2) direct $N$-body integration for situations when the secular approach breaks down, using the algorithmic chain regularization code \textsc{MSTAR} \citep{2020MNRAS.492.4131R}. Due to stellar evolution or other processes, the secular approach can break down in an initially stable hierarchical system when the system becomes dynamically unstable on short timescales \citep[e.g.,][]{2012ApJ...760...99P,2021arXiv210713620H,2021arXiv210804272T}, or when the orbit-averaging approximation is no longer justified \citep[e.g.,][]{2012ApJ...757...27A,2014ApJ...781...45A,2016MNRAS.458.3060L,2018MNRAS.481.4907G,2018MNRAS.481.4602L,2019MNRAS.490.4756L,2020MNRAS.494.5492H}. The \mse~code dynamically switches between the two integration methods, thereby taking advantage of the speed of the secular approach when it is valid, and the accuracy of direct $N$-body integration when the secular approximation breaks down.

In both methods (1) and (2), post-Newtonian correction terms are taken into account. In the secular case, tidal evolution is also included under the assumption of equilibrium tides \citep{1981A&A....99..126H,1998ApJ...499..853E}, with the efficiency of tidal dissipation computed using the prescription of \citet{2002MNRAS.329..897H}.

\subsection{Stellar evolution}
\label{sect:meth:stel}
Rather than modelling stellar evolution with 1D or higher-dimensional codes which are numerically expensive, \mse~treats stellar evolution by incorporating the fast analytic fitting formulae to detailed stellar evolution models from \citet{2000MNRAS.315..543H}, i.e., \textsc{SSE}. These formulae also include mass loss due to stellar winds, and spin-down due to magnetic braking \citep{1983ApJ...275..713R}. We take into account the effects of wind mass loss on the orbits by assuming mass loss occurs adiabatically, i.e., $m_\mathrm{enc} a_i$ and $e_i$ are constant, with $m_\mathrm{enc}$ is the enclosed mass, and $a_i$ and $e_i$ the orbital semimajor axis and eccentricity, respectively. 

When neutron stars (NSs) or black holes (BHs) are formed following a supernova (SN) explosion, \mse~takes into account the mass lost during the SN event assuming that the mass is removed instantaneously from the system. Natal kicks are taken into account by sampling from a Maxwellian distribution with dispersion $\sigma_{\mathrm{kick}} = 265\,\kms$ for NSs \citep{2005MNRAS.360..974H} and $\sigma_{\mathrm{kick}} = 50\,\kms$ for BHs (this corresponds to `kick distribution model 1' from \citealt{2021MNRAS.502.4479H}; we consider other kick distributions to be beyond the scope of this paper).

\subsection{Binary interactions}
\label{sect:meth:bin}
A number of `binary interactions', i.e., interactions between two stars, are taken into account in \mse~(see \S\ref{sect:meth:tr} below for `triple' interactions). These include mass transfer between two stars directly orbiting each other (i.e., in the case of a triple, mass transfer between the inner binary components). Here, details such as the mass transfer rate, stellar aging/rejuvenation, and the conditions for unstable transfer (i.e., whether or not CE evolution will ensue) are modelled according to \citet{2002MNRAS.329..897H}. One significant difference between \mse~and \citet{2002MNRAS.329..897H} is that \citet{2002MNRAS.329..897H} assumes that mass transfer occurs in circular orbits, whereas \mse~adopts the analytic model of \citet{2019ApJ...872..119H} to describe the orbital response to mass transfer in eccentric orbits. The latter can be important in multiple systems, e.g., when secular evolution is important \citep[e.g.,][]{2020A&A...640A..16T}.

In \mse, when the donor is a giant star and the mass ratio between the donor and accretor is sufficiently large (see \citealt{2002MNRAS.329..897H} for details) and/or the estimated mass transfer timescale would be shorter than the donor's dynamical timescale or the orbital period, RLOF is assumed to lead to unstable mass transfer, i.e., CE evolution. The latter is taken into account adopting the $\alpha$-CE prescription, with $\alpha_\mathrm{CE}=1$ (the default value in \mse). Other details of the outcomes of CE evolution are also adopted similarly from \citet{2002MNRAS.329..897H}. 

Accretion of material from stellar winds onto companions can be important in close orbits. This process is taken into account in \mse~by adopting the Bondi-Hoyle-Lyttleton formalism \citep{1939PCPS...35..405H,1944MNRAS.104..273B}. We currently only allow for wind accretion between two objects directly orbiting each other, i.e., we do not model wind accretion of the tertiary star by wind mass in the inner binary, and vice versa.

When stars or compact objects physically collide, \mse~determines the properties of the remnant (its new stellar type, mass, etc.) using similar assumptions as those of \citet{2002MNRAS.329..897H}. For example, two merging MS stars are assumed to result in a single MS star (with no mass loss, as assumed by \citealt{2002MNRAS.329..897H}).

\subsection{Triple interactions}
\label{sect:meth:tr}
In systems with more than two objects, it is possible that a star fills its Roche lobe around a binary component, rather than a single object. \mse~checks for RLOF onto such extended objects. For the RLOF criterion, it assumes --- for simplicity --- the standard \citet{1983ApJ...268..368E} fit of the Roche lobe radius, with the companion interpreted as a point mass; the extended nature of the companion is not taken into account (see \citealt{2020MNRAS.491..495D} for a detailed study into the latter). Using the same criterion for stable transfer in the binary case (cf. \S\ref{sect:meth:bin}), \mse~determines whether the resulting mass transfer onto the inner binary is stable or unstable. The mass transfer rate is determined using the prescription of \citet{2002MNRAS.329..897H}, treating the companion as a single object. The subsequent long-term evolution of triple RLOF is modelled by adopting a number of simplified prescriptions, motivated by more detailed simulations \citep{2014MNRAS.438.1909D,2020MNRAS.498.2957C,2021MNRAS.500.1921G}. 

\begin{table*}
\begin{center}
\begin{tabular}{ccccccccccccccc}
\toprule 
\S & $m_1$ & $m_2$ & $m_3$ & $a_1$ & $a_2$ & $e_1$ & $e_2$ & $i_1$ & $i_2$ & $\omega_1$ & $\omega_2$ & $\Omega_1$ & $\Omega_2$ & $i_\mathrm{rel}$ \\
\midrule
\ref{sect:ex:1} & 2.532 & 1.279 & 3.639 & 0.186 & 7.227 & 0.734 & 0.589 & 1.167 & 2.107 & 2.402 & 5.448 & 3.404 & 3.259 & 0.950 \\
\ref{sect:ex:3} & 1.671 & 1.561 & 2.925 & 0.431 & 3.859 & 0.107 & 0.174 & 2.145 & 0.709 & 3.638 & 1.776 & 4.819 & 2.196 & 2.661 \\
\ref{sect:ex:4} & 3.853 & 0.631 & 4.212 & 0.923 & 8.339 & 0.417 & 0.491 & 2.658 & 2.256 & 4.556 & 4.040 & 1.999 & 0.940 & 0.743 \\
\ref{sect:ex:5} & 11.778 & 5.279 & 16.529 & 1.799 & 8.632 & 0.808 & 0.201 & 0.464 & 2.226 & 0.708 & 2.184 & 3.563 & 5.813 & 2.446 \\
\bottomrule
\end{tabular}
\end{center}
\caption{Initial conditions for the examples presented in \S\ref{sect:ex}. Masses $m_i$ are in units of $\msun$, and semimajor axes $a_i$ are in units of $\au$. The orbital angles (inclinations $i_i$, arguments of periapsis $\omega_i$, and longitudes of the ascending node $\Omega_i$) are expressed in radians. }
\label{table:exICs}
\end{table*}

\subsubsection{Stable triple mass transfer}
\label{sect:meth:tr:stab}
In the case of stable mass transfer, we invoke `circumstellar triple mass transfer', i.e., mass transfer of a tertiary star onto an inner binary. The amount of mass accreted onto the inner binary is assumed to depend on whether or not an accretion disk can form around the inner binary. The latter is determined using the results of \citet{1976ApJ...206..509U}. If an accretion disk forms, we assume that the inner binary can efficiently accrete material from the tertiary star. The accretion rates onto the inner binary components are then 
\begin{align}
\dot{m}_1 &= \alpha_{\mathrm{TMT,\,disk,\,1}} \cdot \dot{m}_3; \\
\dot{m}_2 &= \alpha_{\mathrm{TMT,\,disk,\,2}} \cdot \dot{m}_3,
\end{align}
where $m_1$ and $m_2$ are the inner binary primary and secondary masses\footnote{The inner binary primary is defined as the currently most massive star in the inner binary.}, $m_3$ is the donor's mass, and the two parameters are set in this work to $\alpha_{\mathrm{TMT,\,disk,\,1}} = 0.6$, and $\alpha_{\mathrm{TMT,\,disk,\,2}} = 0.3$. When no accretion disk forms, mass accretion by the inner binary is assumed to be inefficient \citep[e.g.,][]{2014MNRAS.438.1909D}, and we adopt
\begin{align}
\dot{m}_1 &= \alpha_{\mathrm{TMT,\,no\,disk,\,1}} \cdot \dot{m}_3; \\
\dot{m}_2 &= \alpha_{\mathrm{TMT,\,no\,disk,\,2}} \cdot \dot{m}_3,
\end{align}
where we set $\alpha_{\mathrm{TMT,\,no\,disk,\,1}} = 0.1$, and $\alpha_{\mathrm{TMT,\,no\,disk,\,2}} = 0.1$.

The precise values of the accretion $\alpha$ parameters described above are somewhat arbitrary; they are highly uncertain and depend on the configuration. Detailed simulations are required to determine them accurately. We justify our approach by noting that, in most cases, the amount of mass transferred during stable triple mass transfer in our population synthesis simulations is very small (cf. \S\ref{sect:popsyn:stab}). Therefore, changing the $\alpha$ parameters will not qualitatively affect our results. 

During stable triple mass transfer, the inner orbit is assumed to evolve in a manner that can be described by successive CE-like events, as found by \citet{2014MNRAS.438.1909D}. Here, we adopt $(\alpha \lambda)_{\mathrm{TMT}} = 5$. Any mass not accreted by the inner binary is assumed to leave the inner binary in the form of an adiabatic wind (which affects the outer orbit, but without wind mass accretion by the tertiary star, and not affecting the inner orbit). To describe the outer orbital response to the mass transferred, \mse~also follows \citet{2014MNRAS.438.1909D} by adopting standard expressions for the orbital response to non-conservative mass transfer in circular orbits \citep[e.g.,][]{1997A&A...327..620S}. For more details, we refer to \citet{2021MNRAS.502.4479H}. 

In the above treatment, we do not take into account of any effects that an accretion disk around the inner binary (if formed due to triple mass transfer) could have. Specifically, viscous torques from such a disk will lead to angular momentum-transfer and hence impact the long-term evolution of the inner binary, especially if the latter is eccentric \citep[e.g.,][]{2017MNRAS.466.1170M,2019ApJ...871...84M,2020ApJ...889..114M}. Depending on parameters such as the binary eccentricity, the inner orbit might contract or expand -- in the former case, the inner binary could be driven to merge, whereas in the latter case dynamical instability with the tertiary could be triggered.

\subsubsection{Unstable triple mass transfer}
\label{sect:meth:tr:unstab}
Unstable RLOF of a tertiary star onto an inner binary is assumed in \mse~to lead to triple CE evolution, during which the inner binary becomes engulfed by the tertiary star's envelope. For isolated binary stars, CE evolution still is highly uncertain, since it is generally not clear how the envelope can be efficiently ejected (see, e.g., \citealt{2013A&ARv..21...59I} for a review). Evidently, the triple case is even more uncertain, since there are few observational constraints (although, see \citealt{2019MNRAS.484.4711M}), and since it involves hydrodynamic and gravitational dynamical processes which generally require detailed modelling. Simulations carried out so far \citep[e.g.,][]{2020MNRAS.498.2957C,2021MNRAS.500.1921G} show that a variety of outcomes is possible. The binary can merge  inside the tertiary's envelope relatively quickly, or a dynamical instability can ensue. During the latter, as the inner binary moves closer to the tertiary's dense core, strong gravitational scattering can occur between the inner binary components and the core of the tertiary star, similarly to three-body scattering in dense stellar systems \citep[e.g.,][]{1983ApJ...268..319H}, except that scattering in this case occurs in a gaseous environment. During these instabilities, components can potentially be ejected from the system, or collisions can occur. 

Detailed simulations of triple CE are beyond the capabilities of population synthesis codes such as \mse. To nevertheless model this phase in at least some qualitative manner, \mse~adopts a simplified scheme that is motivated by detailed simulations. Specifically, we use the standard $\alpha$-CE prescription, applied to the tertiary star and the outer orbit, to determine the new tertiary stellar properties and the new outer orbit after CE ejection (assumed to be successful), taking the inner binary to be a point mass. Further assuming that the inner orbit remains unaffected, the triple (sub)system with the new outer orbit is evaluated for dynamical stability (according to the criterion of \citealt{2001MNRAS.321..398M}). If the new outer orbit would be unstable, the tertiary's remnant core is instead `parked' at the boundary between stability and instability, and \mse~switches to direct $N$-body integration to determine the outcome of the chaotic interaction involving the inner binary stars and the donor's core\footnote{During the chaotic phase in \mse, other stars in the system (not applicable to triples) are also taken into account in the direct $N$-body integration, although they are necessarily widely separated from the triple system undergoing CE evolution, and hence will likely have little effect.}. Here, the the $N$-body system is initiated based on the current mean anomalies, which are advanced in time in the code according to the mean motions. 

Evidently, this approach neglects all hydrodynamical effects during the triple CE process. However, it is able to capture some of the effects observed in detailed simulations (e.g., \citealt{2021MNRAS.500.1921G}), such as collisions or ejections of components during the triple CE event. We remark that our approach is consistent with the prescription of \citet{2020MNRAS.498.2957C}. 

We refer to \S\ref{sect:dis:cav} for a further discussion on the caveats of our approach to model triple RLOF evolution.

\section{Examples}
\label{sect:ex}

In this section, we illustrate some of the channels involving triple RLOF as found in our population synthesis calculations (cf. \S\ref{sect:popsyn}). The initial conditions for all systems discussed are given in Table~\ref{table:exICs}. For each example, we show mobile diagrams \citep{1968QJRAS...9..388E} representing the system at key points during the evolution (up to a system age of $10\,\gyr$). The (\textsc{SSE}) stellar types in the mobile diagrams are indicated with colors (refer to the legend at the top of each example figure). The arrows emanating from stars (where applicable, namely when the code is in direct $N$-body integration mode) indicate the velocity directions, i.e., the unit velocity vectors projected onto the $(x,y)$-plane. Red arrows connecting two stars indicate strong interactions such as RLOF, CE evolution, and physical collisions.

\subsection{Triple CE inner binary merger}
\label{sect:ex:1}

\begin{figure*}
\iftoggle{ApJFigs}{
\includegraphics[width=1.\linewidth,trim = 40mm 105mm 0mm 0mm]{system_2166_mobile}
}{
\includegraphics[width=1.\linewidth,trim = 40mm 105mm 0mm 0mm]{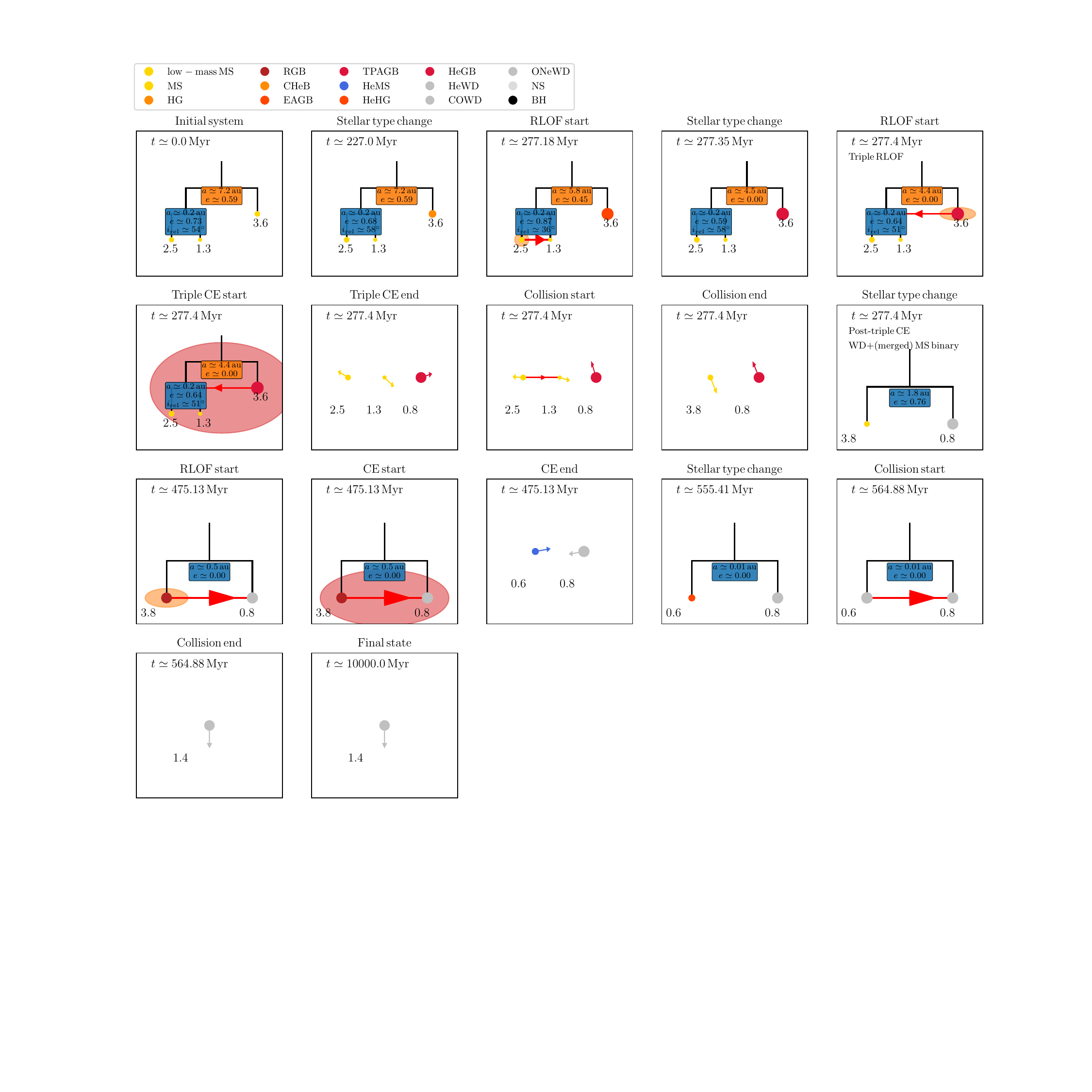}
}
\caption{Mobile diagrams for a system in which unstable tertiary RLOF is followed by a merger in the inner binary system. The merged star, a new MS star, forms a tight binary with the tertiary star (now a WD). The latter evolves into a double WD system that merges into a single WD (the total WD mass before merger is just below the Chandrasekhar mass; a SNe Ia explosion is  assumed not to occur). When applicable, the current mutual inclination between the inner and outer orbits ($i_\mathrm{rel}$) is indicated. Note that the mobiles (black solid lines) are only shown for bound (sub)systems and when the code is in secular integration mode. Directly following (triple) CE, the code switches to direct $N$-body integration, in which case the orbital configuration may not be stable; stars are in this case depicted as in isolation (although they may be actually bound).}
\label{fig:ex1}
\end{figure*}

\F~\ref{fig:ex1} shows an example of a triple system in which the tertiary star is the most massive and fills its Roche lobe around the inner binary. Mass transfer is assumed by \mse~to be unstable; the new triple system is dynamically unstable and the subsequent evolution is modelled with direct $N$-body integration, quickly leading to a merger of the two MS stars in the inner binary. 

An eccentric MS+WD binary (with the WD being the remnant of the tertiary star) remains at $\sim 277\,\myr$. The merged MS star evolves to a giant star within $\sim 200\,\myr$ and this invokes CE evolution, producing a tight circular CO WD+CO WD binary that quickly merges. The total mass of the two merging WDs, $\simeq 1.40\,\msun$, is slightly below the assumed threshold in \mse~for a Type Ia SNe (SNe Ia) explosion (i.e., the Chandrasekhar mass, $\simeq 1.44\,\msun$), hence the two WDs are assumed to merge quietly into a single (highly massive) CO WD. We note that this assumption is somewhat inaccurate as such a scenario has been suggested to lead to either accretion-induced collapse \citep[e.g.,][]{2015MNRAS.453.1910S,2019MNRAS.484..698R,2020RAA....20..135W,2020MNRAS.494.3422L} or (sub-luminous) sub-Chandrasekhar-mass SNe \citep[e.g.,][]{2010ApJ...714L..52S,2012ApJ...747L..10P,2013ApJ...770L...8P,2021A&A...649A.155G}. Therefore, some of our systems in this channel may produce a transient phenomenon instead of a CO WD.

\subsection{Triple CE binary+single outcome}
\label{sect:ex:3}

\begin{figure*}
\iftoggle{ApJFigs}{
\includegraphics[width=0.8\linewidth,trim = 30mm 95mm 0mm 0mm]{system_1885_mobile}
}{
\includegraphics[width=0.8\linewidth,trim = 30mm 95mm 0mm 0mm]{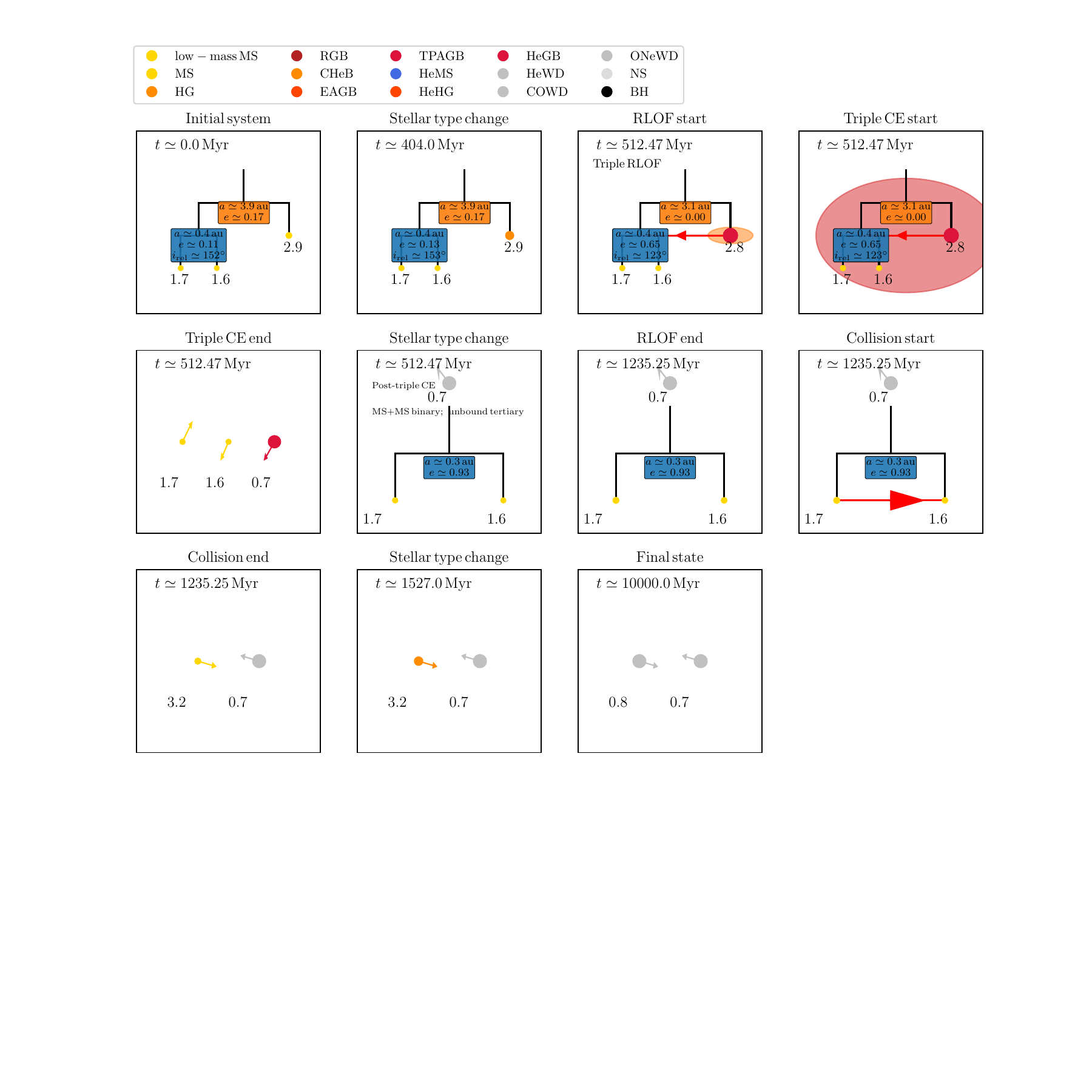}
}
\caption{Mobile diagrams for a system in which triple CE leads to the unbinding of the tertiary star. The inner binary (consisting of two MS stars) survives initially, but its orbital periapsis decreases. It merges at a later time when the stars expand during the MS, forming a new single MS star that eventually becomes a CO WD.}
\label{fig:ex3}
\end{figure*}

\F~\ref{fig:ex3} showcases another outcome of triple CE evolution, namely the unbinding of the tertiary star, leaving the inner binary as a bound system. Triple CE is triggered at $t\sim 512\,\myr$; before this time, the inner binary was mildly excited in eccentricity due to secular evolution. Triple CE is unstable and leads to three-body interactions; the tertiary star --- now a CO WD --- escapes, whereas the remaining binary has a smaller semimajor axis and higher eccentricity than before the triple CE phase. The two MS stars in the perturbed inner binary are close to merging, with radii of $\simeq 0.0077\,\au$ and $\simeq 0.0074\,\au$, respectively (sum $\simeq 0.015\,\au$), whereas the orbital periapsis distance is $\simeq 0.021\,\au$. 

As the two MS star evolve and grow in radius, they finally merge at $t\sim 1235\,\myr$, producing a new MS star. The latter ultimately becomes a CO WD.

\subsection{Triple CE binary+single outcome with exchange and SNe Ia explosion}
\label{sect:ex:4}

\begin{figure*}
\iftoggle{ApJFigs}{
\includegraphics[width=0.8\linewidth,trim = 30mm 30mm 0mm 5mm]{system_9284_mobile}
}{
\includegraphics[width=0.8\linewidth,trim = 30mm 30mm 0mm 5mm]{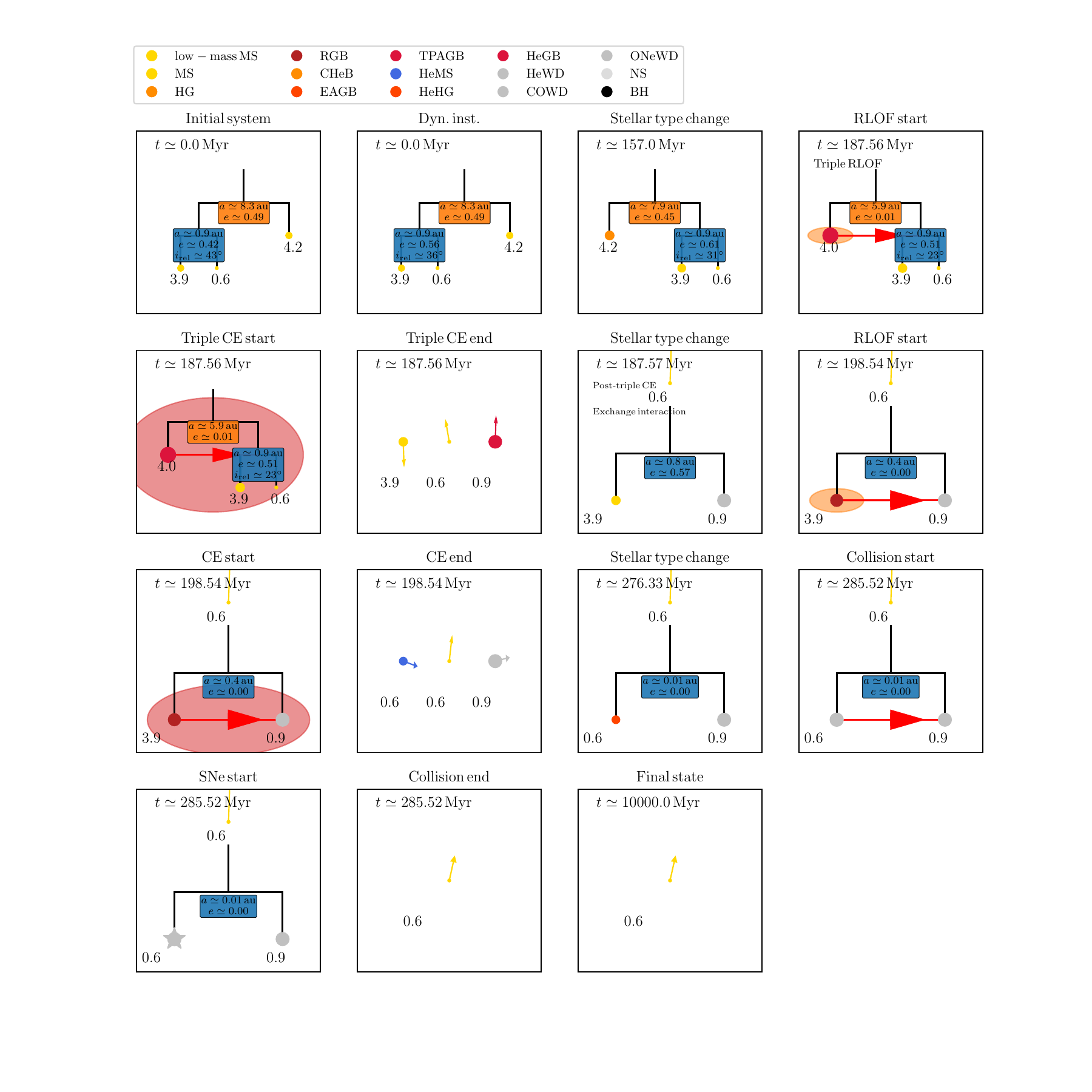}
}
\caption{Mobile diagrams for a system in which triple CE occurs with an exchange interaction, leaving a binary with swapped components. The latter subsequently leads to a merger of two WDs and a SNe Ia explosion. Note that the original inner binary, in isolation, would not have been conducive to ultimately producing a SNe explosion. }
\label{fig:ex4}
\end{figure*}

In this example (\F~\ref{fig:ex4}), triple CE leads to a binary+single configuration, similarly as in \S\ref{sect:ex:3}. However, in this case, an exchange interaction occurs during which one of the inner binary stars is ejected, whereas a new, tight binary forms at $\simeq 188\,\myr$, consisting of the other inner binary star and the tertiary star, by this time a CO WD. Such an exchange interaction is plausible, given that the escaping star is significantly lower in mass ($m\simeq 0.6\,\msun$) compared to the other two stars (each $\sim 4 \, \msun$). 

The MS star in the remaining binary evolves and triggers CE evolution with its CO WD companion when it becomes an RGB star at $t \sim 198\,\myr$. This strips the envelope of the giant star, producing a He star+CO WD binary in a tight circular orbit. The He star evolves into another CO WD. The double CO WD system subsequently merges; the combined WD mass exceeds the Chandrasekhar mass, and \mse~assumes that a SNe Ia explossion occurs, leaving no remnant. The only remaining object from the triple at $t=10\,\gyr$ is the low-mass MS star that was ejected during the triple CE phase.

\subsection{Stable triple mass transfer}
\label{sect:ex:5}

\begin{figure*}
\iftoggle{ApJFigs}{
\includegraphics[width=0.8\linewidth,trim = 30mm 30mm 0mm 5mm]{system_731_mobile}
}{
\includegraphics[width=0.8\linewidth,trim = 30mm 30mm 0mm 5mm]{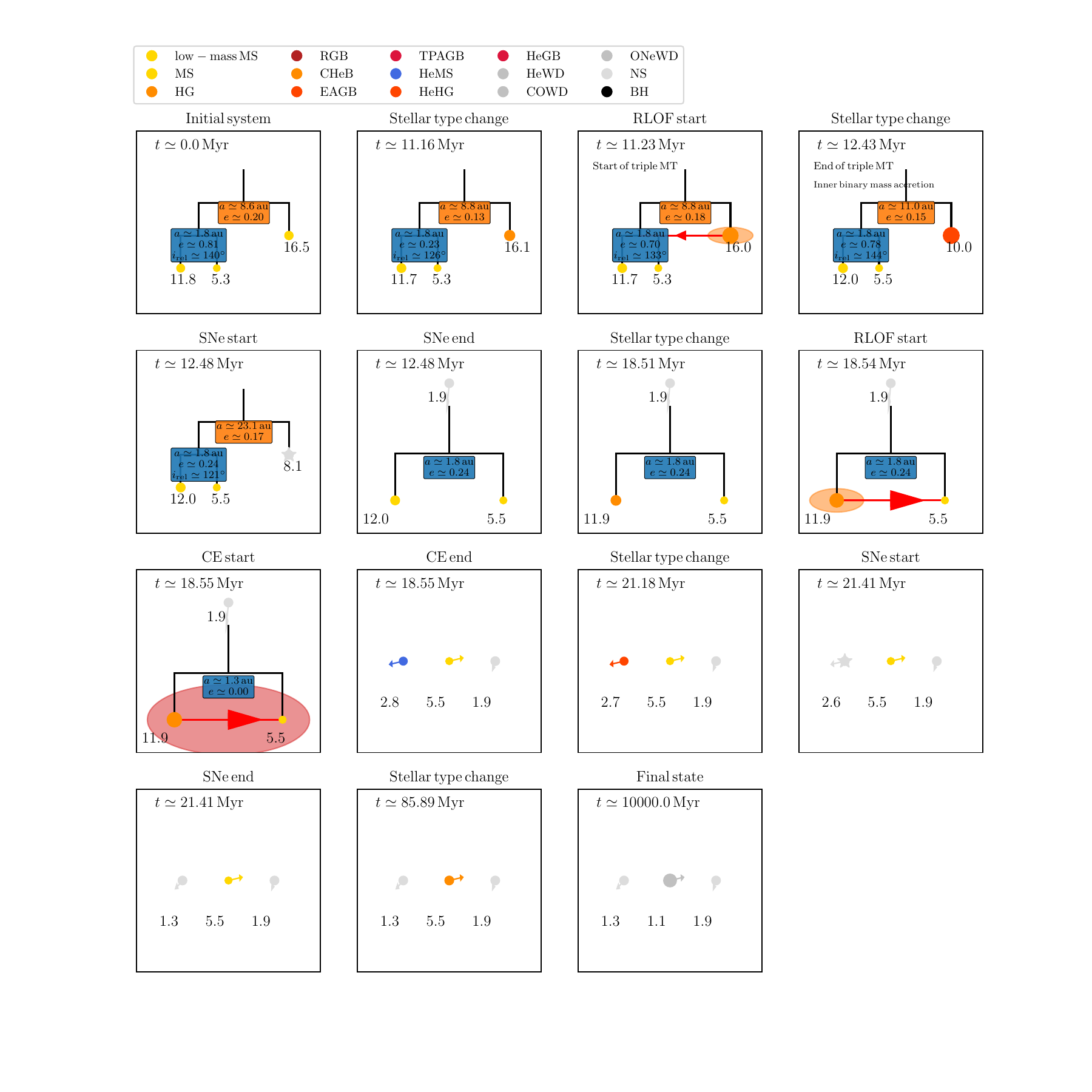}
}
\caption{Mobile diagrams for a system in which stable triple mass transfer occurs, and a significant amount of mass is accreted onto the inner binary.}
\label{fig:ex5}
\end{figure*}

Our last example (\F~\ref{fig:ex5}) illustrates a significant amount of mass transfer from the tertiary star onto the inner binary in a stable manner (we remark that a significant mass transfer amount is actually atypical for stable triple mass transfer, cf. \S\ref{sect:popsyn:stab}). The tertiary star, relatively massive with $m_3\simeq 16.5\,\msun$, fills its Roche lobe around the inner binary at $t\sim 11\,\myr$. Within approximately 1 Myr, it transfers some of its mass to the inner binary, whereas also continuing to lose mass in the form of stellar winds. During this time, the inner binary components increase in mass by $\sim 0.3$ and $\sim 0.2\,\msun$, respectively (note that the inner stars are also losing mass in the form of winds; both effects are taken into account). The outer orbit widens significantly during this process, and triple mass transfer stops. At $t\sim 12\,\myr$, the tertiary star undergoes SNe and becomes a NS; the sudden mass loss and natal kick unbind it from the inner binary system.

The inner binary system survives for another $\sim 10 \, \myr$, until the most massive star becomes a giant and triggers CE evolution. As it later undergoes a SNe explosion, the remaining binary becomes unbound. Its companion, still a MS star, eventually becomes a CO WD. The final state at $t=10\,\gyr$ consists of three unbound compact objects.

\section{Simulation setup}
\label{sect:ICs}
Here, we describe the assumptions made to generate a population of triples, the implied normalization, and the simulation settings that underlie the results discussed in \S\ref{sect:popsyn}. Since only a small fraction of all triple systems are expected to undergo triple RLOF (see \S\ref{sect:popsyn:frac}), we restrict the parameter space considered in order to increase the number of systems undergoing triple RLOF in our simulations.

\subsection{Monte Carlo sampling}
\label{sect:ICs:mc}
In our sampling procedure, we generate triple systems from assumed underlying distributions, during which we reject a system it if does not satisfy the conditions of being initially dynamically stable, and of the stars not initially filling their Roche lobes. We consider this set to be the `complete' (synthetic) triple population. Subsequently, we consider a subset of the complete population, namely those systems that we expect are more likely to lead to triple RLOF evolution. This subset constitutes a fraction $f_\mathrm{calc}$ of the complete population. 

\subsubsection{Complete population}
\label{sect:ICs:mc:com}
To sample the complete population, we assume an underlying initial mass function (IMF) for the primary star of \citet{2001MNRAS.322..231K,2002Sci...295...82K}, i.e., a broken power law with
\begin{align}
\label{eq:imf}
\frac{\mathrm{d}N}{\mathrm{d} m_1} \propto \left \{ 
\begin{array}{cc} 
m_1^{-1.3}, & 0.08\,\msun<m_1\leq0.5 \, \msun; \\
m_1^{-2.3}, & 0.5\,\msun<m_1\leq1\,\msun; \\
m_1^{-2.35}, & 1\,\msun<m_1\leq100\,\msun.
\end{array}
\right.
\end{align}
We define the primary star to be the most massive star in the inner binary system. We subsequently sample an inner binary mass ratio $q_1$ from a flat distribution ($0<q_1\leq1$), and compute the inner binary secondary mass according to $m_2 = q_1 m_1$. The tertiary star mass, $m_3$ is computed by sampling another uniform mass ratio $q_2$ ($0<q_2\leq 1$), such that $q_2 = m_3/(m_1+m_2)$. We reject the sampled masses if any mass is less than $0.08\,\msun$, or larger than $100\,\msun$. Short-period binaries prefer more equal masses \citep[e.g.,][]{2000A&A...360..997T,2003A&A...397..159H,2006ApJ...639L..67P,2017ApJS..230...15M} whereas we assume a flat inner orbital mass ratio distribution. However, we do not expect our results on the onset of triple RLOF (i.e., evolution prior to triple RLOF) to be strongly dependent on the assumed inner mass ratio distribution; the latter affects in particular the strength of the secular octupole-order terms \citep[e.g.,][]{2011ApJ...742...94L}, but secular evolution turns out to be relatively unimportant for the evolution prior to triple RLOF (see \S\ref{sect:popsyn:trlof:ICs} below). The outcome of triple RLOF, in particular triple CE evolution, could on the other hand be affected by the inner binary mass ratio distribution. For example, if the inner binary has highly unequal-mass components and triple CE evolution leads to a dynamical instability phase, then this will more likely lead to the ejection of the lighter inner binary component. Nevertheless, we consider a detailed investigation into the dependence on the inner binary mass ratio to be beyond the scope of the paper (see also \S\ref{sect:ICs:mc:set} below in relation to computational constraints). 

We remark that our approach allows for the tertiary star to be more massive than the primary star in the inner binary system, although not more massive than the total inner binary mass (i.e., we assume $q_2\leq 1$). In the Multiple Star Catalogue (MSC; \citealt{1997A&AS..124...75T,2018ApJS..235....6T}), the fraction of triples with $q_2 > 1$ is $\simeq 0.07$, indicating that the latter population is not negligible. However, it should be noted that the MSC is not volume-complete and hence subject to observational biases. Nevertheless, because of our restriction $q_2\leq 1$, our results will likely yields lower limits to the occurrence of triple RLOF. 

Subsequently, we sample the inner ($i=1$) and outer ($i=2$) orbital semimajor axes $a_i$, adopting a lognormal distribution in the orbital period if the primary stellar mass is $m_1<3\,\msun$ \citep{1991A&A...248..485D,2010ApJS..190....1R}, and flat in log $a_i$ if $m_1 > 3 \, \msun$ \citep{2007ApJ...670..747K}. We note that this is qualitatively consistent with \citet{2017ApJS..230...15M}. The semimajor axis range is $10^{-3} \, \au<a_i<10^5\,\au$. Orbits wider than $\sim 10^5\,\au$ are dissociated in the Galactic potential \citep[e.g.,][]{2018MNRAS.474.4412F}. The lower limit, on the other hand, is to ensure the tightest possible (inner) systems; in practice, the tightest orbit is set by the requirement of not initially Roche lobe filling, meaning that $a_1\gtrsim 10^{-2} \, \au$.

To sample the orbital eccentricities $e_i$, we adopt a Gaussian distribution for both inner and outer orbital eccentricities with a mean of $\mu=0.4$ and standard deviation of $\sigma=0.4$, as suggested by \citet{2013ARA&A..51..269D} as a reasonable approximation to observed multiple systems.

The orbital angles (inclinations $i_i$, arguments of periapsis $\omega_i$, and longitudes of the ascending node $\Omega_i$, for the inner and outer orbits) are sampled to be consistent with random orbital orientations. For simplicity and lack of detailed observational constraints, we ignore the observational fact that tight triples tend to be more coplanar \citep{2017ApJ...844..103T}.

A sampled system as described above is rejected if it is initially unstable according to the analytic stability criterion of \citet{2001MNRAS.321..398M}. Furthermore, we reject a system if its inner orbit has a sufficiently small periapsis distance such that the inner binary components would initially be Roche lobe filling, using the fit of \citet{1983ApJ...268..368E} for the Roche lobe radius, and estimating the stellar radii (for the purposes of sampling only) as $R_\star = R_\odot \, (m_i/\msun)^{0.7}$ \citep[e.g.,][]{1994sse..book.....K}. We remark that the stability and Roche lobe filling criteria affect the orbital distributions (semimajor axes and eccentricities) in particular. 

\subsubsection{Subset of systems}
\label{sect:ICs:mc:sub}
From the `complete' set of triples sampled as described in \S\ref{sect:ICs:mc:com}, we select those which are more likely to lead to triple RLOF evolution compared to the average system. These constraints were motivated based on earlier sets of simulations without parameter space restrictions.

Specifically, we select triples with the constraints $m_3>1\,\msun$, $m_3>m_1$, and $a_2 < 10^3\,\au$. The first constraint ensures (approximately) that the tertiary star will undergo a stellar type change within $10\,\gyr$, the second that the tertiary star will evolve fastest, and the third constraint implies that only relatively tight triples are included. These limits are verified a-posteriori in \S\ref{sect:popsyn:trlof:ICs}.

With these restrictions and assumptions, we find that the subset of systems evolved with \mse~constitutes a fraction $f_\mathrm{calc} \simeq 0.04220$ of the complete set of triples. We refer to Appendix~\ref{app:norm} for a description on how fractions observed in the simulations are converted into occurrence rates. 

\subsection{Simulation settings}
\label{sect:ICs:mc:set}
We simulate with \mse~each system from the sampled subset (\S\ref{sect:ICs:mc:sub}) for a duration of $t_\mathrm{end}=10\,\gyr$. \mse~includes the effects of fly-bys in the field and these are enabled in the simulations (using the default settings; see \citealt{2021MNRAS.502.4479H}). However, we remark that, given that the systems considered here are relatively tight triple systems with outer orbits $\leq 10^3\,\au$, fly-bys are negligible. 

The fact that many systems in our subset are tight triples means that they are computationally expensive to integrate for an age of $10\,\gyr$. For example, a triple with an inner orbital period of 10 d and an outer orbital period of 100 d has a ZLK timescale \citep[e.g.,][]{2015MNRAS.452.3610A} of $t_\mathrm{ZLK} \sim (100/10) \,100\,\mathrm{d} \simeq 2.7 \,\yr$, which is tiny compared to $10 \, \gyr$, and implies that a very large number of secular oscillations needs to be computed. This is exacerbated by the short additional apsidal motion timescale in the inner orbit (due to either tidal bulges or general relativity). Given the large computational expense, we impose a maximum CPU wall time per system of 20 hr. The latter was exceeded for $\approx 7\%$ of sampled systems, but we do not expect this to significantly affect our results. Furthermore, in light of computational constraints, we limit this work to a single population of triples, and do not explore the dependence of our results on model assumptions such as the CE $\alpha$ parameter.

\section{Statistical results}
\label{sect:popsyn}
Here, we present the main results of the population synthesis. We first consider some general fractions and rates (\S\ref{sect:popsyn:frac}), and then focus on conditions for systems to undergo triple RLOF (\S\ref{sect:popsyn:trlof}), those systems for which triple RLOF leads to CE evolution (\S\ref{sect:popsyn:trce}), and those systems for which triple RLOF proceeds stably (\S\ref{sect:popsyn:stab}). 

\subsection{Fractions and rates}
\label{sect:popsyn:frac}

A summary of our population synthesis results in tabular form is given in Tables~\ref{table:fractions} and \ref{table:rates}. Table~\ref{table:fractions} shows the fractions (in per cent) of the occurrence of triple RLOF relative to all triples (i.e., the `complete' population, cf. \S\ref{sect:ICs:mc:com}; this takes into account the fraction $f_\mathrm{calc}$ as described in \S\ref{sect:ICs:mc:sub}), subdivided into stable and unstable RLOF. For the unstable case, branching fractions are quoted for triple CE outcomes (relative to all triple CE cases). Also included are fractions for different types of merger outcomes in the case of triple CE (relative to all triple CE merger cases). Arrows ($\rightarrow$) indicate subsets of the above channel, with the relative fraction (in per cent) in this channel indicated in brackets. 

In the case of triple CE, we define the `outcome' by considering the state of the system within $1\,\myr$ of the time of onset of triple CE. This is motivated by the fact that triple CE, if producing an unstable triple, is expected to lead a chaotic three-body interaction that will typically last for a short time. We remark that setting a significantly later delay time for this analysis would imply that further evolution (such as binary interactions, if a binary system survives) could be important, whereas our goal here is to focus exclusively on the `immediate' outcome of triple CE. 

\begin{table}
\begin{tabular}{ll}
\toprule 
Description & $f(\%)$ \\
\midrule
Tertiary RLOF (relative to all triples) \\
\midrule
Tertiary\,RLOF & $0.058 \pm 0.002$ \\
$\rightarrow$ Stable\,mass\,transfer \, (64.0\%) & $0.037 \pm 0.001$ \\
$\rightarrow$ Unstable\,(triple\,CE) \, (36.0\%) & $0.021 \pm 0.001$ \\
\midrule
Triple CE outcomes (general) \\
\midrule
Merger(s) & $76.3 \pm 0.4$ \\
Binary+Single & $22.9 \pm 0.2$ \\
$\rightarrow$ No\,Exchange (98.4\%) & $22.6 \pm 0.2$ \\
$\rightarrow$ Exchange (1.61\%) & $0.37 \pm 0.03$ \\
Triple & $0.74 \pm 0.04$ \\
\midrule
Triple CE outcomes (mergers) \\
\midrule
Inner\,Merger & $60.5 \pm 0.4$ \\
Multiple\,Mergers & $35.6 \pm 0.3$ \\
Inner+Tertiary\,Merger & $3.9 \pm 0.1$ \\
\bottomrule
\end{tabular}
\caption{Fractions (in per cent) of the occurrence of triple RLOF relative to all triples, subdivided into stable and unstable RLOF. For the unstable case, branching fractions are shown for triple CE outcomes (relative to all triple CE cases). Also shown are fractions for different types of merger outcomes in the case of triple CE (relative to all triple CE merger cases). Arrows ($\rightarrow$) indicate subsets of the above channel, with the relative fraction (in per cent) in this channel indicated in brackets. }
\label{table:fractions}
\end{table}

\begin{table}
\begin{tabular}{ll}
\toprule 
Description & $R/(10^{-4} \, \yr^{-1})$ \\
\midrule
All\,Mergers\,Involving\,TRLOF (100.0 \%) & $1.197 \pm 0.025$ \\
$\rightarrow$ MS-MS (18.8 \%) & $0.225 \pm 0.011$ \\
$\rightarrow$ MS-Giant (53.2 \%) & $0.637 \pm 0.018$ \\
$\rightarrow$ MS-WD (6.1 \%) & $0.073 \pm 0.006$ \\
$\rightarrow$ MS-BH/NS (0.1 \%) & $0.002 \pm 0.001$ \\
$\rightarrow$ Giant-Giant (1.2 \%) & $0.014 \pm 0.003$ \\
$\rightarrow$ Giant-WD (17.6 \%) & $0.210 \pm 0.011$ \\
$\rightarrow$ Giant-BH/NS (0.2 \%) & $0.002 \pm 0.001$ \\
$\rightarrow$ WD-WD (2.7 \%) & $0.033 \pm 0.004$ \\
$\rightarrow$ WD-BH/NS (0.0 \%) & $0.001 \pm 0.001$ \\
$\rightarrow$ BH/NS-BH/NS (0.1 \%) & $0.001 \pm 0.001$ \\
\bottomrule
\end{tabular}
\caption{Galactic rates of mergers in our simulations that involve triple RLOF (both stable and unstable triple mass transfer). The first row gives the total Galactic rate, whereas other rows (indicated with arrows) give rates for several combinations of stellar types. Here, MS stars include both centrally H and He-burning stars. Numbers in round brackets give the percentage contribution of each stellar type combination relative to all mergers involving triple RLOF.}
\label{table:rates}
\end{table}

As seen in Table~\ref{table:fractions}, the overall triple mass transfer fraction is small, i.e., $\simeq 0.06\%$ (we remind the reader that with `triple mass transfer', we consider exclusively the case when the tertiary star overflows its Roche lobe around the inner binary). Of these systems, $\simeq 64\%$ lead to stable transfer, whereas $\simeq 36\%$ is unstable and expected to lead to triple CE evolution. We note that `stable' mass transfer includes any system in which triple RLOF occurs, i.e., the amount of mass transferred can be negligible (cf. \S\ref{sect:popsyn:stab}). 

When triple CE occurs, the most likely outcome is one or multiple mergers (i.e., several mergers within $1\,\myr$). Most commonly, a single merger occurs in the inner binary system. Alternatively, multiple mergers occur; e.g., the inner binary merges first and the remnant star subsequently merges with the tertiary star, or the tertiary star merges with an inner binary star and the merger product subsequently merges with the inner binary companion star. In a small fraction of cases ($\simeq 4\%$), a single merger occurs between the tertiary star and one of the inner binary stars. Triple CE mergers are discussed in more detail in \S\ref{sect:popsyn:trce}. 

The next likely triple CE outcome is the formation of a binary system with a third unbound star. In the majority of this case of binary+single formation, the tertiary star becomes unbound from the inner binary. In a small fraction of cases ($\simeq 2\%)$, an exchange interaction occurs and the tertiary star forms a new binary with one of the inner binary components, whereas the other inner binary component is ejected. A stable triple remains only in a small fraction of cases ($\simeq 1 \%$). Generally, these fractions show that triple CE is very likely ($\simeq 99\%$) to disrupt the hierarchical nature of the triple system (although, see the caveats as discussed in \S\ref{sect:dis:cav}). 

Table~\ref{table:rates} quotes Galactic event rates for all systems undergoing triple RLOF and with one or multiple mergers occurring subsequently (see Appendix~\ref{app:norm} for the normalization calculation). The combined rate (not making a distinction between stellar type of the merging stars involved), is $\simeq 1.2 \times 10^{-4} \, \yr^{-1}$. Interestingly, this rate is comparable to the rate of mergers in triples following dynamical instability \citep{2012ApJ...760...99P,2021arXiv210713620H,2021arXiv210804272T}. Although the overall fraction of dynamically unstable systems ($\sim 0.5\%$ relative to all triples, see, e.g., \citealt{2021arXiv210713620H}) is larger than the fraction of systems in which triple RLOF occurs ($\simeq 0.06\%$), the latter is more likely to lead to mergers. For reference, the observed Galactic rate \citep{2014MNRAS.443.1319K} is $\sim 0.5 \, \yr^{-1}$. Therefore, both triple dynamical instability as well as triple RLOF are not expected to contribute significantly to the Galactic merger rate.

\begin{figure*}
\iftoggle{ApJFigs}{
\includegraphics[width=0.5\linewidth]{a1s}
\includegraphics[width=0.5\linewidth]{a2s}
\includegraphics[width=0.5\linewidth]{q2s}
\includegraphics[width=0.5\linewidth]{i_rels}
}{
\includegraphics[width=0.5\linewidth]{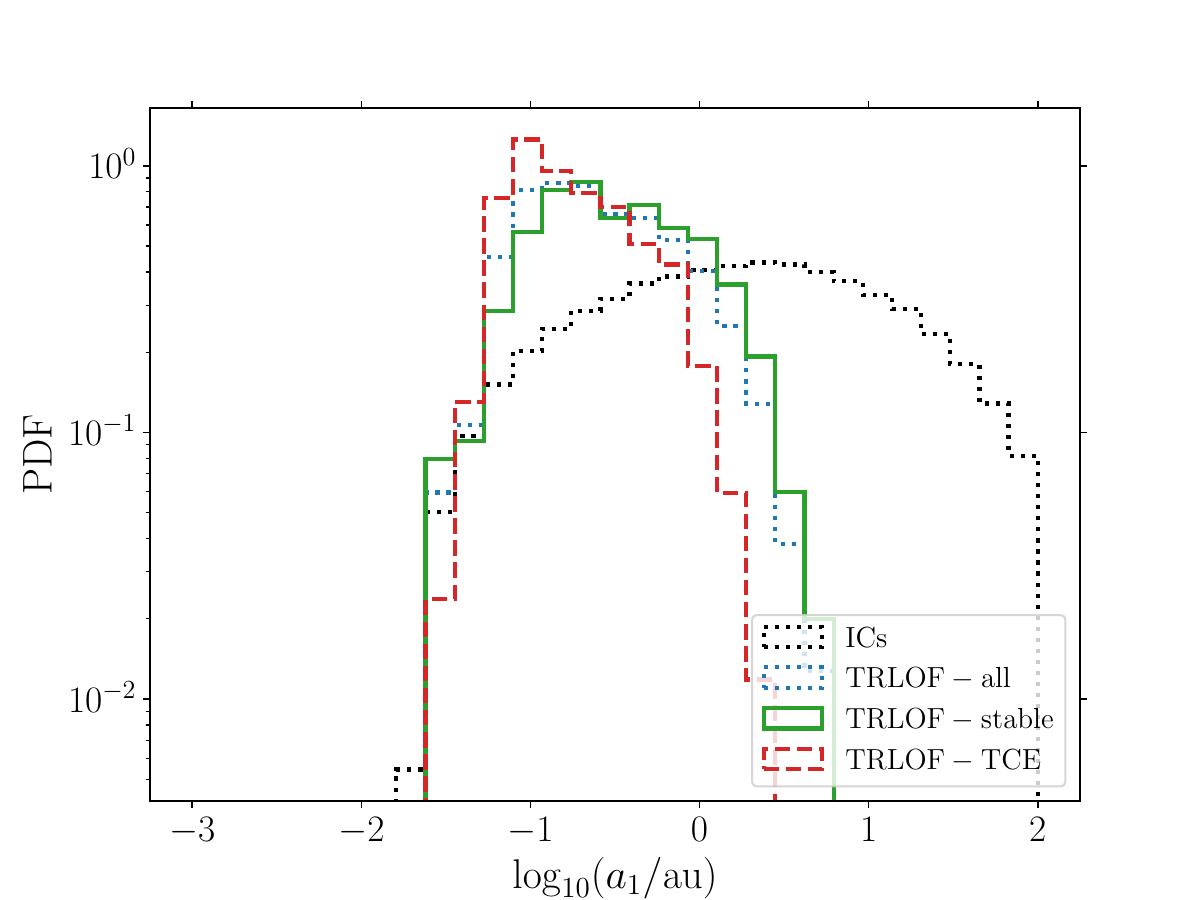}
\includegraphics[width=0.5\linewidth]{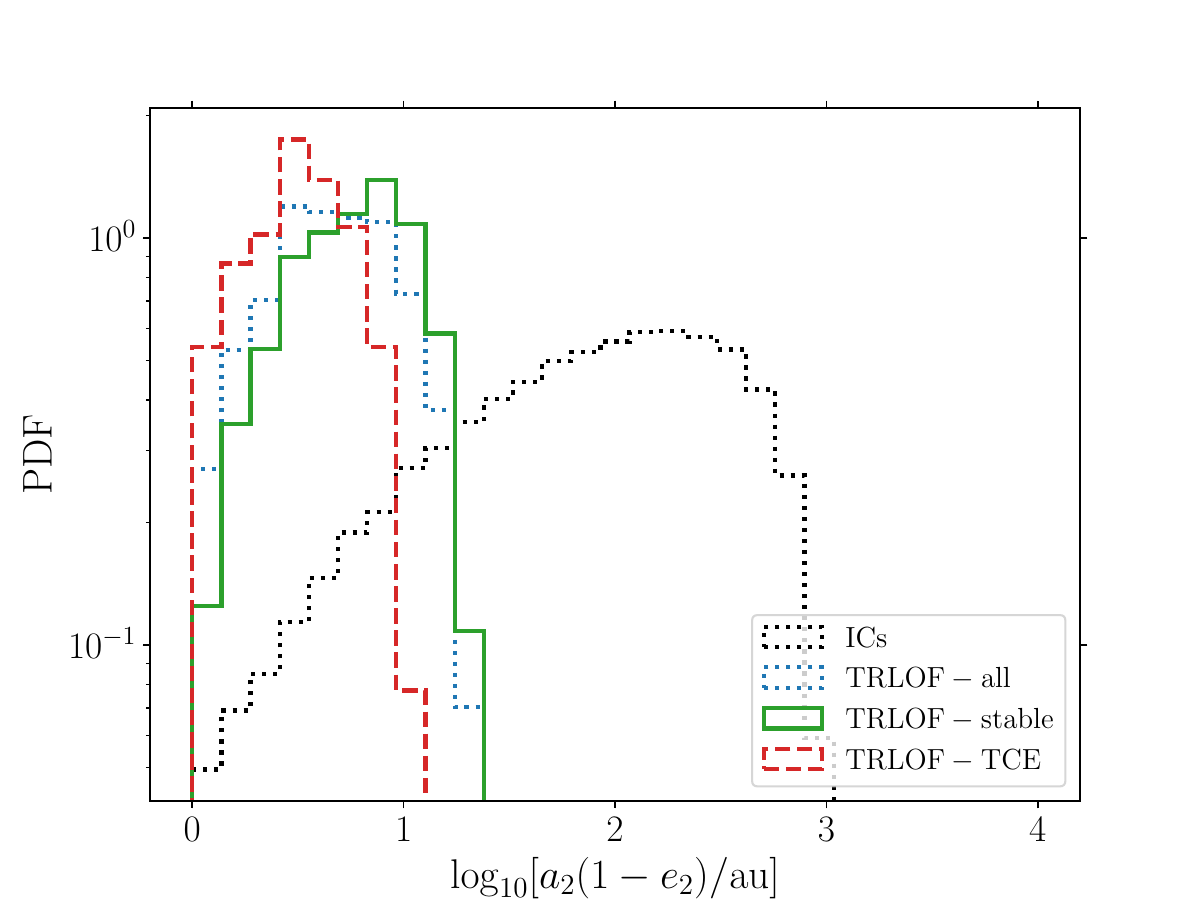}
\includegraphics[width=0.5\linewidth]{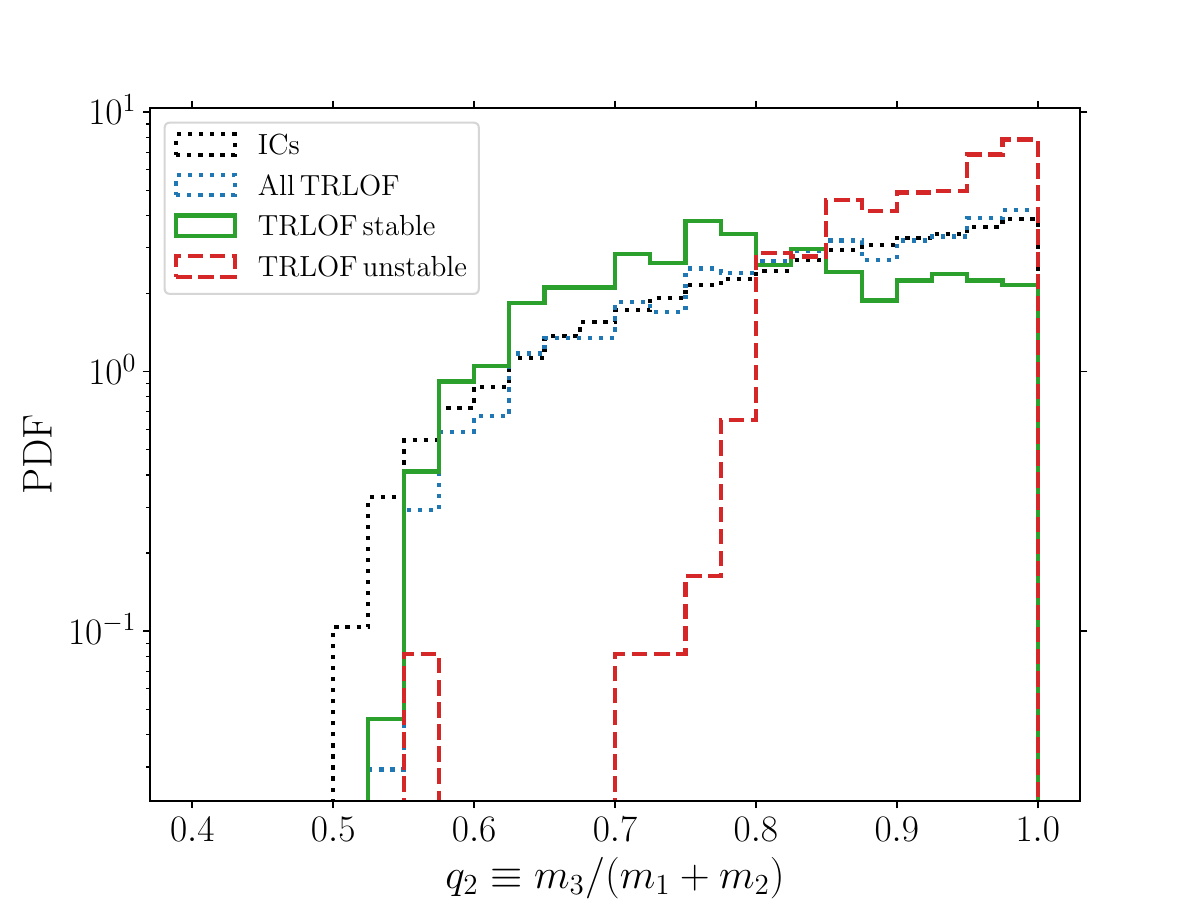}
\includegraphics[width=0.5\linewidth]{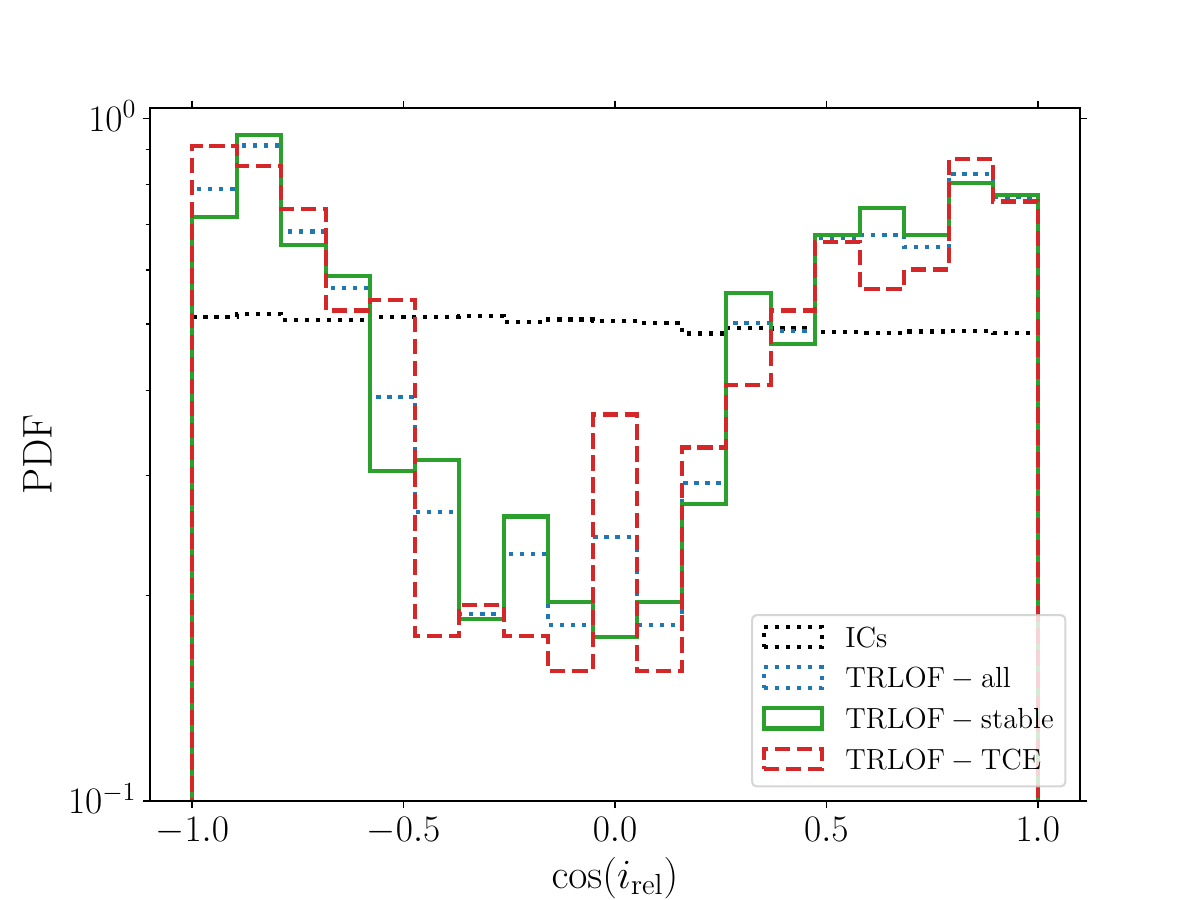}
}
\caption{Probability density distributions for initial values of several parameters, making a distinction between all systems considered in the population synthesis calculations (black dotted lines), all systems undergoing triple RLOF (blue dotted lines), triple RLOF systems leading to stable transfer (solid green lines), and triple RLOF systems leading to CE evolution (red dashed lines). The four panels show initial distributions of the inner semimajor axis, $a_1$, the outer orbit periapsis distance, $a_2(1-e_2$), the outer orbital mass ratio $q_2 \equiv m_3/(m_1+m_2)$, and the cosine of the mutual inclination, $i_\mathrm{rel}$.  }
\label{fig:trlof_ICs}
\end{figure*}

\subsection{Systems leading to triple RLOF}
\label{sect:popsyn:trlof}
Here, we consider in more detail the initial conditions of systems undergoing triple RLOF, and their evolution leading up to this stage. 

\subsubsection{Initial conditions}
\label{sect:popsyn:trlof:ICs}
In \F~\ref{fig:trlof_ICs}, we show probability density distributions of the initial values for several parameters, making a distinction between all systems included in the population synthesis calculations (black dotted lines; we remind the reader that these systems comprise a subset of all triples, cf. \S\ref{sect:ICs:mc:sub}), all systems undergoing triple RLOF (blue dotted lines), triple RLOF systems leading to stable transfer (solid green lines), and triple RLOF systems leading to CE evolution (red dashed lines). The four panels show initial distributions of the inner semimajor axis, $a_1$, the outer orbit periapsis distance, $a_2(1-e_2$), the outer orbital mass ratio $q_2 \equiv m_3/(m_1+m_2)$, and the cosine of the mutual inclination ($i_\mathrm{rel}$).  

All systems undergoing triple RLOF show significantly smaller $a_1$ and $a_2(1-e_2)$ than the range included in the population synthesis simulations, justifying the imposed cutoff in $a_2$ (cf. \S\ref{sect:ICs:mc:sub}). With regard to $a_1$, there exists a slight preference for larger inner orbital semimajor axes for stable mass transfer, and similarly (and more pronounced) for $a_2(1-e_2)$. The differences between stable and unstable triple RLOF are clearly revealed in $q_2$ --- stable systems typically have smaller outer mass ratios than unstable ones, as expected given the prescription implemented in \mse~(cf. \S\ref{sect:meth:bin}). As a caveat, however, we remark that the latter prescription is highly uncertain, even more so in triples, and --- unfortunately --- many interesting astrophysical implications depend on the understanding of the stability of mass transfer \citep[e.g.,][]{2021arXiv210705702G}. 

The bottom-right panel in \F~\ref{fig:trlof_ICs} shows an interesting aspect of systems undergoing triple RLOF: whereas the initial population was assumed to have a distribution consistent with random orbital orientations (i.e., a flat distribution in $\cos i_\mathrm{rel}$), systems undergoing triple RLOF show a clear lack of systems with $i_\mathrm{rel}$ initially near $90^\circ$. Systems that are initially mutually highly inclined are more likely to undergo an inner binary merger due to secular evolution, meaning that a system undergoing triple RLOF must typically have relatively low initial mutual inclination. Furthermore, an asymmetry is apparent, with fewer triple RLOF systems having retrograde orientations ($\cos i_\mathrm{rel}<0$). This can be attributed to the fact that secular excitation is typically more efficient for retrograde orientations, even at lowest-order approximation as long as the test-particle assumption is relaxed \citep[e.g.,][]{2021MNRAS.500.3481H}.

\begin{figure}
\iftoggle{ApJFigs}{
\includegraphics[width=\linewidth]{times}
}{
\includegraphics[width=\linewidth]{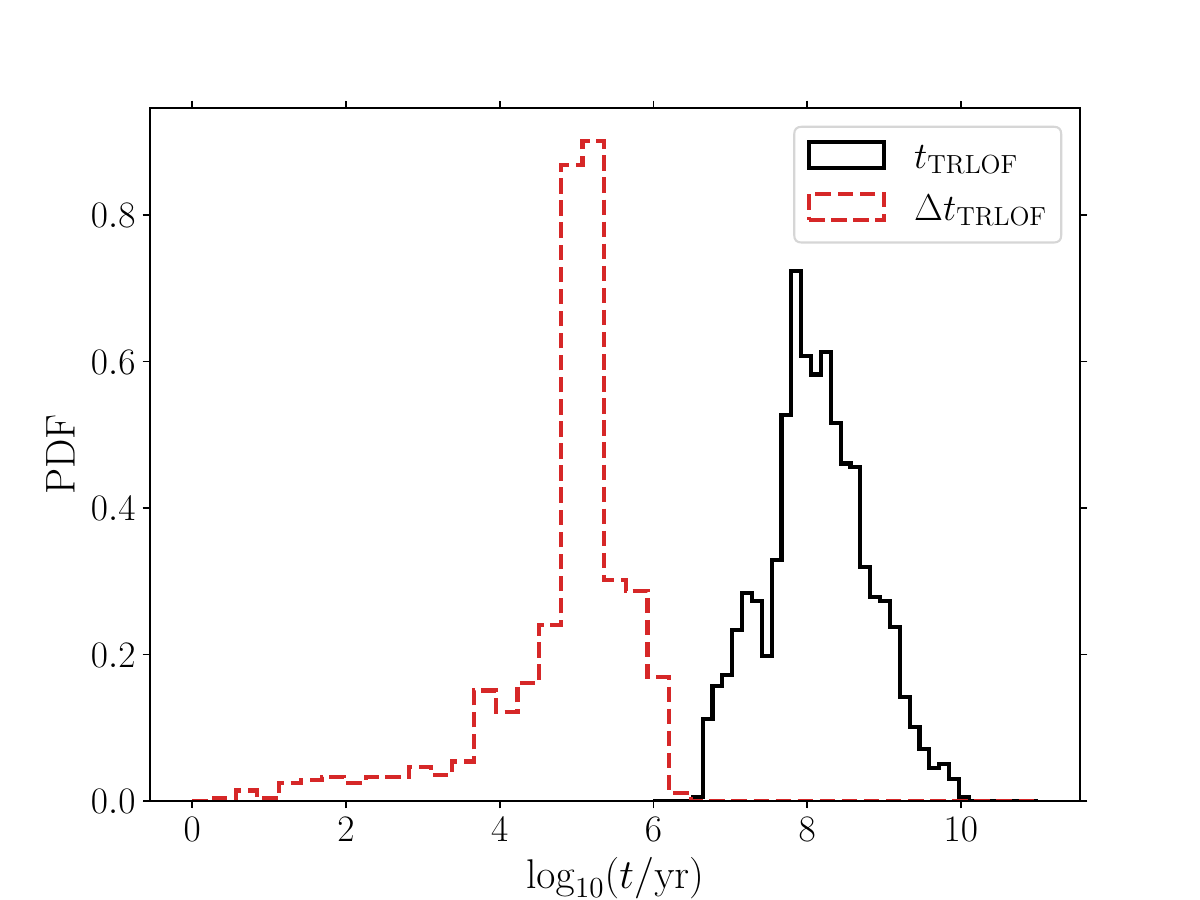}
}
\caption{Distribution of the age of the system in the Monte Carlo simulations when triple RLOF was triggered (black solid line). The red dashed line shows the distribution of the duration of triple RLOF when mass transfer is stable (here, multiple RLOF episodes are possible for a single system).}
\label{fig:trlof_times}
\end{figure}

\begin{figure}
\iftoggle{ApJFigs}{
\includegraphics[width=1.0\linewidth,trim = 10mm 0mm 20mm 0mm]{final_TRLOF_merger_sts}
}{
\includegraphics[width=1.0\linewidth,trim = 10mm 0mm 20mm 0mm]{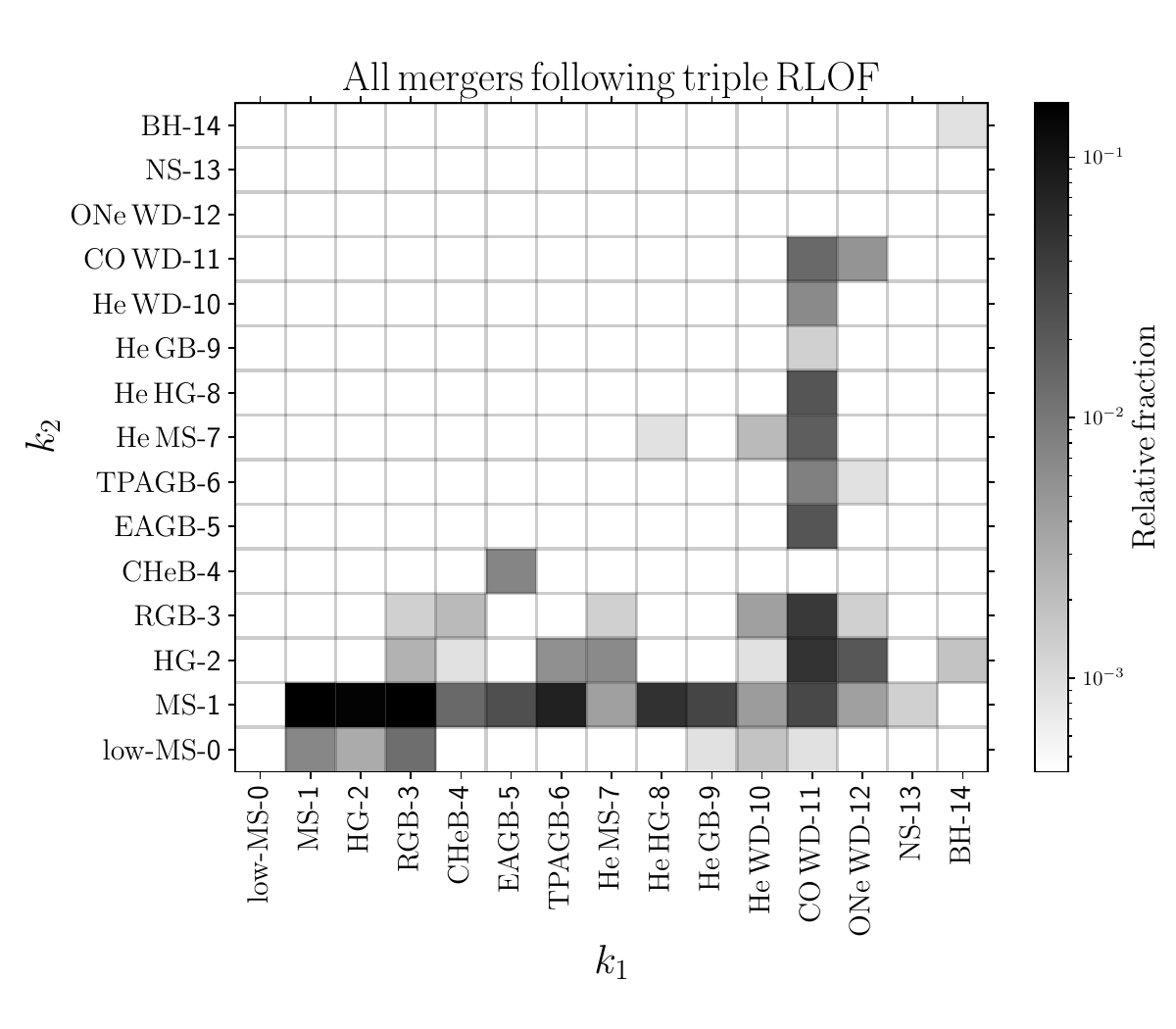}
}
\caption{Stellar types of merging stars for a mergers following any type of triple RLOF. Greyscale shading indicates the fraction of a particular combination of stellar types (see \citealt{2000MNRAS.315..543H}), relative to all mergers involving triple RLOF. }
\label{fig:trlof_merger_sts}
\end{figure}

\begin{figure}
\iftoggle{ApJFigs}{
\includegraphics[width=\linewidth]{TRLOF_a_changes}
\includegraphics[width=\linewidth]{TRLOF_e_changes}
\includegraphics[width=\linewidth]{TRLOF_m_changes}
}{
\includegraphics[width=\linewidth]{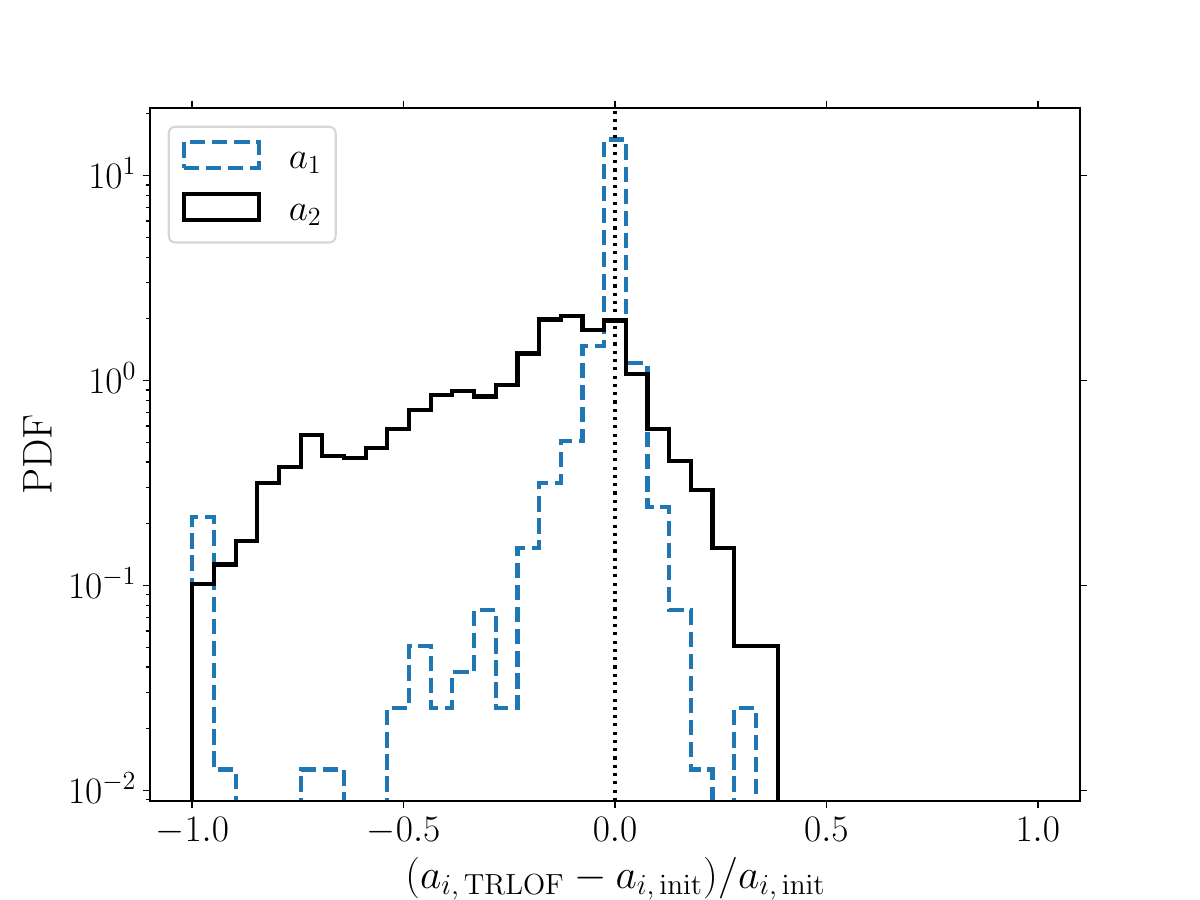}
\includegraphics[width=\linewidth]{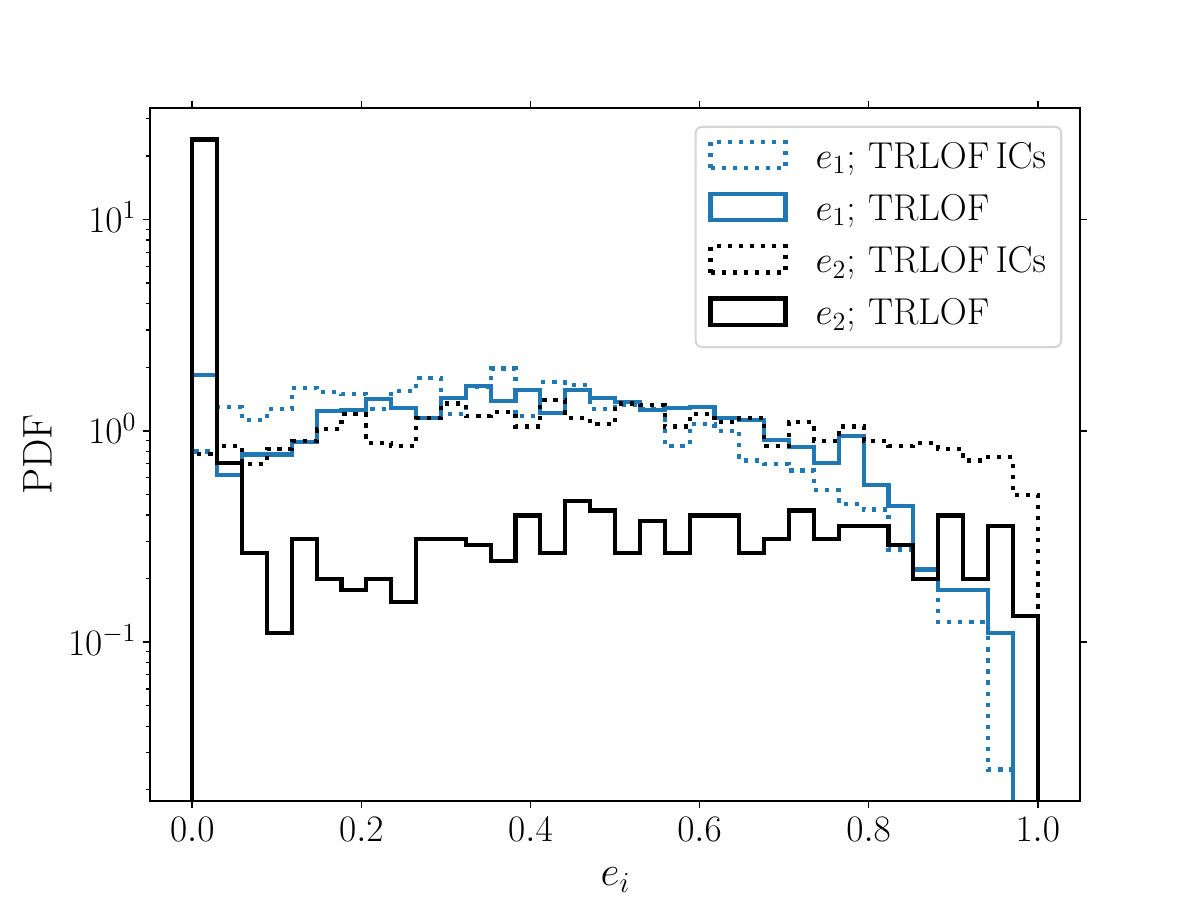}
\includegraphics[width=\linewidth]{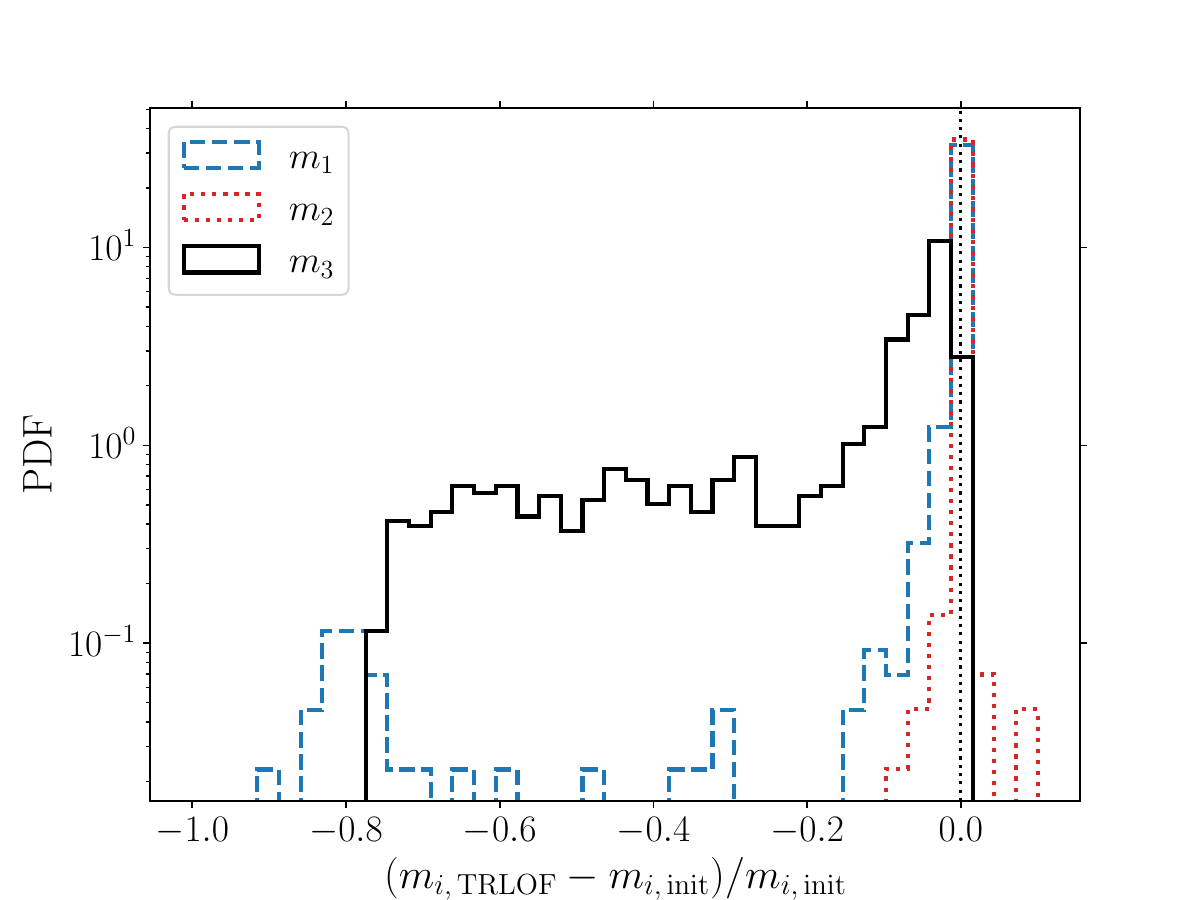}
}
\caption{Changes in the triple system between formation and the onset of triple RLOF. The top panel shows the fractional changes in $a_1$ and $a_2$ from birth until triple RLOF. The middle panel shows distributions of the inner and outer eccentricities for systems undergoing triple RLOF, i.e., their initial distributions (dotted lines), as well as their distributions at the moment of triple RLOF (solid lines). The bottom panel shows fractional changes in the three masses between birth and the beginning of triple RLOF.  }
\label{fig:trlof_changes}
\end{figure}

\subsubsection{Evolution until triple RLOF}
\label{sect:popsyn:trlof:ev}
\F~\ref{fig:trlof_times} (black line) shows the distribution of the system ages in the Monte Carlo simulations when triple RLOF was triggered. These ages are mainly set by the evolutionary timescale of the tertiary star, since the latter must expand in order to fill its Roche lobe around the inner binary. For systems undergoing triple RLOF, the distribution of $m_3$ peaks near $6\,\msun$, and the (MS) lifetime of such a star is $\sim 10\,\gyr\,(m_3/\msun)^{-2.8} \sim 66\,\myr$ \citep[e.g.,][]{1994sse..book.....K}. This is consistent with the peak near $10^8\,\yr$ in \F~\ref{fig:trlof_times}. The earliest moment of triple RLOF is $\sim 4 \, \myr$, corresponding to the most massive $m_3$, whereas --- in rare cases --- the least massive $m_3$ can still lead to triple RLOF at up to $\sim 10\,\gyr$. 

In \F~\ref{fig:trlof_merger_sts}, we show in greyscale shading relative frequencies of merging systems involving triple RLOF in the stellar type chessboard diagram, i.e., frequency relative to all mergers involving triple RLOF in the $(k_1,k_2)$ plane of stellar types of merging objects (refer to \citealt{2000MNRAS.315..543H} for the detailed stellar type definitions, but in short: 0-1: MS stars; 2, 3, 5, 6: giants; 4: core He burning; 7-9: He stars; 10-12: WDs; 13: NSs; 14: BHs). Here, we include all mergers after the first onset of triple RLOF, i.e., up to a system age of $10 \, \gyr$. The information shown in this figure was also presented in a more condensed form in Table~\ref{table:rates}. 

For systems undergoing triple RLOF, there exists a large range of stellar type combinations. Dominant are MS-MS mergers, as well as giant-MS and giant-WD mergers. WD-WD mergers also occur; these are interesting, e.g., in the context of SNe Ia, and are discussed in more detail in \S\ref{sect:dis:WD}.

In addition to the stellar types of merging stars following triple RLOF, it is of interest to consider the orbital and mass evolution of the system until the onset of triple RLOF. The top panel of \F~\ref{fig:trlof_changes} shows the fractional changes in $a_1$ and $a_2$ from system formation until triple RLOF. These changes can be both positive or negative. It may seem counterintuitive that $a_2$ can also increase --- such expansion can be attributed to wind mass loss of the tertiary star, which tends to widen the outer orbit. However, one should note that, even if $a_2$ is increasing, the tertiary star can still fills its Roche lobe around the inner binary if its radius increases significantly enough. Nevertheless, in a large fraction of cases, the outer orbit shrinks substantially, which can be attributed to strong tidal evolution when the tertiary star becomes a giant star and develops a deep convective envelope. Interestingly, the distribution of the fractional change in $a_2$ peaks near 0, corresponding to no net change in $a_2$.

The inner orbital semimajor typically changes less compared to that of the outer orbit, with the distribution of the fractional change in $a_1$ strongly peaking at 0. Changes in the inner orbit can be attributed to several effects, such as tidal evolution, wind mass loss, and mass transfer (between the inner components). The latter in particular leads to a small peak of fractional changes in $a_1$ near $-1$, corresponding to systems in which the inner binary system has undergone CE evolution before the tertiary star fills its Roche lobe around the inner binary. 

The middle panel of \F~\ref{fig:trlof_changes} shows distributions of the inner and outer eccentricities for systems undergoing triple RLOF, i.e., their initial distributions (dotted lines), as well as their distributions at the moment of triple RLOF (solid lines). For these systems, the outer orbital eccentricity $e_2$ tends to be higher initially than average (i.e., systems not undergoing triple RLOF). By the time of triple RLOF, the outer orbit is typically much more circular, with the distribution of $e_2$ strongly peaking near 0. This can be attributed to tidal evolution. The distribution of $e_1$ also shows some evolution: zero-eccentricity systems are enhanced (due to tidal evolution in the inner orbit), whereas there are more systems with moderate to high eccentricity, i.e., $e_1\gtrsim 0.6$. The latter can be attributed to secular eccentricity excitation in the inner orbit. The effect is only small, however, and can be understood from the bottom-right panel of \F~\ref{fig:trlof_ICs}: in order for a system to become a triple RLOF system, the initial mutual inclination needs to be relatively small, otherwise the inner binary is likely to have merged before triple RLOF could occur. This excludes extreme eccentricity excitation.

The bottom panel of \F~\ref{fig:trlof_changes} shows fractional changes in the three masses between birth and the beginning of triple RLOF. The tertiary star only decreases in mass as a result of wind mass loss and the fractional decrease can be significant, up to $\sim 80\%$. The inner binary components show typically small mass changes by the time of triple RLOF, although the inner binary primary star ($m_1$) shows a local peak at $\sim 80\%$. These correspond to systems in which the inner binary underwent CE before triple RLOF. The inner binary secondary star ($m_2$) shows a fractional increase in some systems, which can be attributed to mass transfer in the inner binary system.

\begin{figure}
\iftoggle{ApJFigs}{
\includegraphics[width=1.0\linewidth,trim = 10mm 0mm 20mm 0mm]{TRCE_merger_stellar_types}
}{
\includegraphics[width=1.0\linewidth,trim = 10mm 0mm 20mm 0mm]{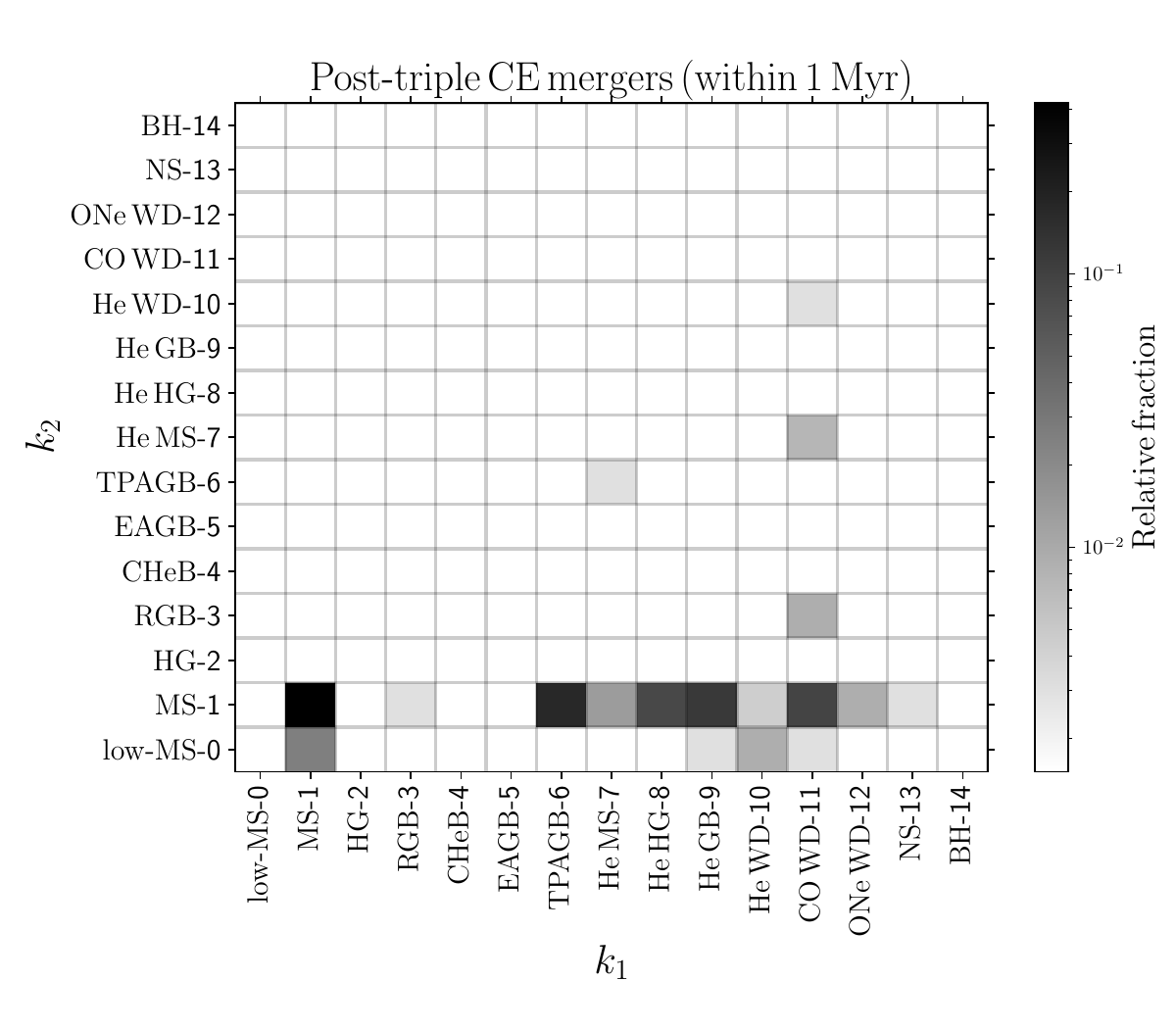}
}
\caption{Stellar types of merging stars following triple CE evolution (within $1\,\myr$ of the onset of triple RLOF). Greyscale indicates the fraction of a particular combination of stellar types (see \citealt{2000MNRAS.315..543H}), relative to all mergers involving triple CE. }
\label{fig:tce_merger_sts}
\end{figure}

\begin{figure}
\iftoggle{ApJFigs}{
\includegraphics[width=\linewidth]{binary_single_properties_a}
\includegraphics[width=\linewidth]{binary_single_properties_e}
}{
\includegraphics[width=\linewidth]{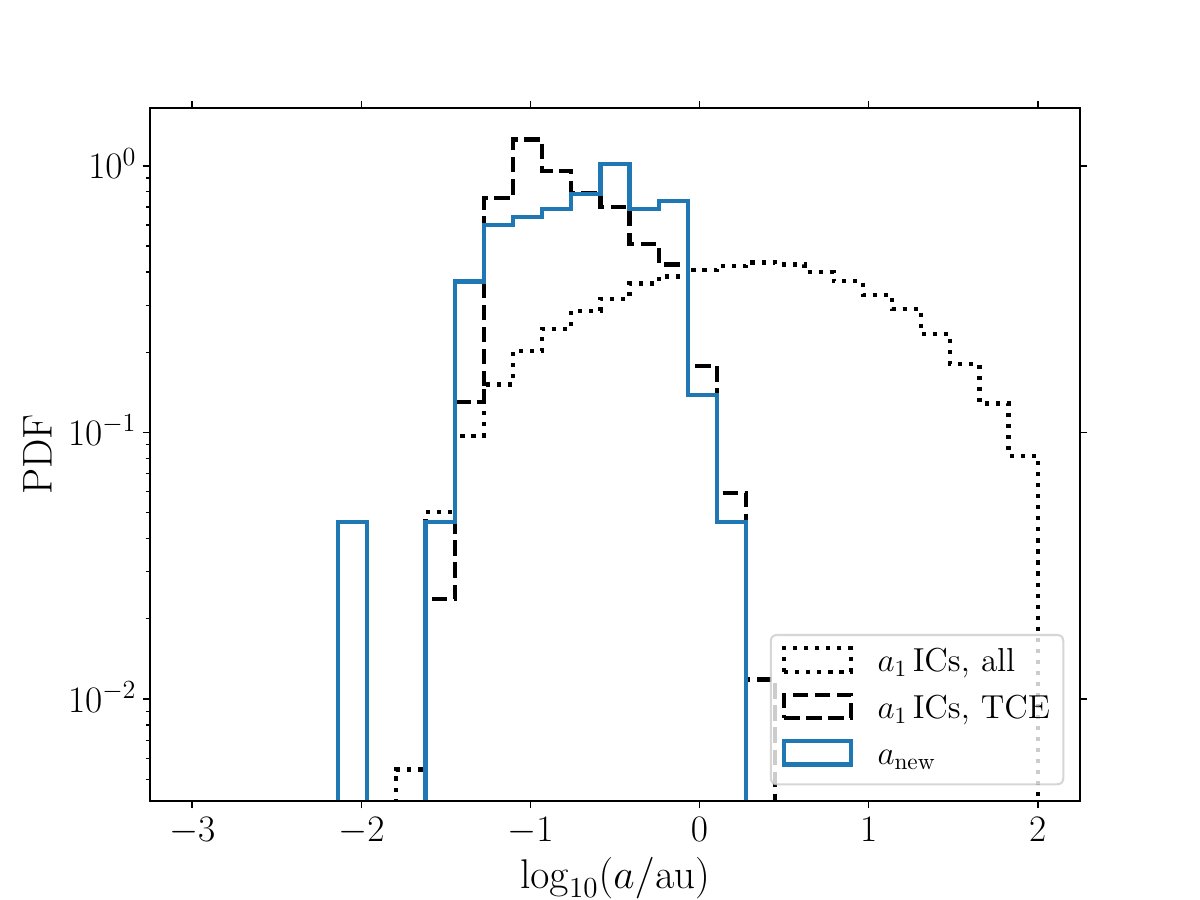}
\includegraphics[width=\linewidth]{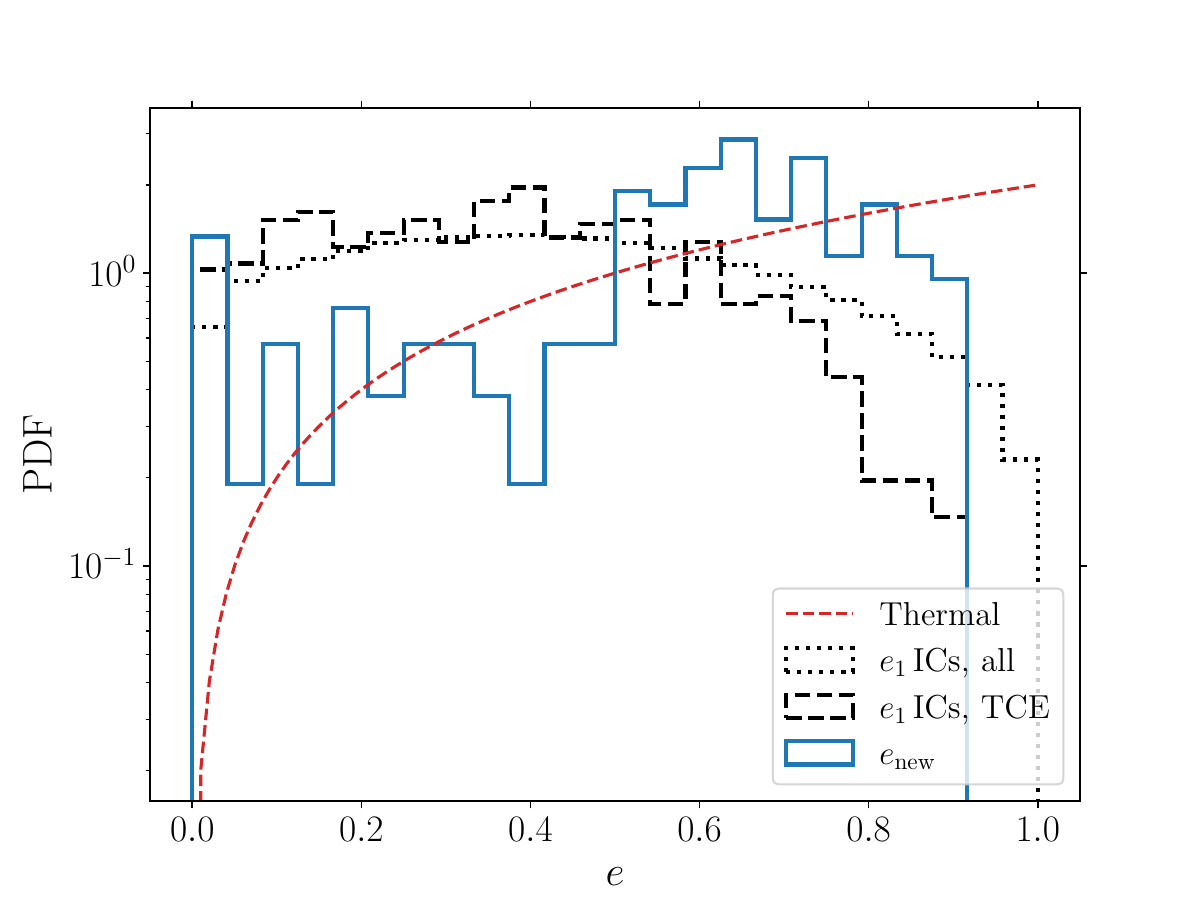}
}
\caption{Distributions of the semimajor axes (top panel) and eccentricities (bottom panel) of the surviving binaries in the case that triple CE leads to a binary+single configuration (within $1\,\myr$). In each panel, shown are the initial distributions for all systems (black dotted lines), the initial distributions for all systems leading to triple CE evolution (black dashed lines), and the distributions of the new binary orbital properties (blue solid lines). In the bottom panel, the red dashed line shows a thermal distribution, $\mathrm{d}N/\mathrm{d} e = 2 e$.}
\label{fig:binary_single}
\end{figure}

\subsection{Triple CE}
\label{sect:popsyn:trce}
Here, we focus on triple CE evolution. \F~\ref{fig:tce_merger_sts} shows a stellar type chessboard plot for merging stars following triple CE evolution (within $1\,\myr$ of the onset of triple RLOF). There are clear differences between these mergers, and all mergers following triple RLOF (associated with both stable and unstable RLOF; cf. \F~\ref{fig:trlof_merger_sts}). Specifically, post triple-CE systems are dominated by inner binary MS-MS mergers, and mergers with more evolved stars are mainly limited to those involving one MS star. In contrast, for all mergers following triple RLOF, although MS-MS comprise a significant fraction, merging objects also typically involve stars which are both more evolved. 

Next, we consider the properties of binaries formed after binary+single formation following triple CE. As discussed in \S\ref{sect:popsyn:frac}, $\simeq 23\%$ of triple CE systems lead to the formation of a binary+single system (within $1\,\myr$). \F~\ref{fig:binary_single} shows distributions of the semimajor axes (top panel) and eccentricities (bottom panel) of the surviving binaries in this scenario; in each panel, given are the initial distributions for all systems (black dotted lines), the initial distributions for all systems leading to triple CE evolution (black dashed lines), and the distributions of the new binary orbital properties (blue solid lines). 

When a new binary is formed after a chaotic three-body interaction, its semimajor axis is similarly distributed compared to the initial inner binary semimajor axis. Comparing the distributions of the initial inner eccentricity and the eccentricity of the formed binary in the case of the binary+single outcome, the latter is significantly more eccentric, showing a larger tail at higher eccentricities, $e \gtrsim 0.6$, and a lack of systems with lower eccentricities. This can be attributed to multiple effects: the inner binary may have become more eccentric by the time of triple RLOF due to secular evolution, and the binary can be excited in eccentricity due to chaotic three-body interactions after triple CE. The latter effect is expected to lead to a thermal eccentricity distribution \citep[e.g.,][]{2016MNRAS.456.4219A,2019ApJ...872..165G}, and the binary eccentricity distribution indeed becomes more consistent with such a distribution (cf. the red dashed line in the bottom panel of \F~\ref{fig:binary_single}, which shows a thermal distribution).

\begin{figure}
\iftoggle{ApJFigs}{
\includegraphics[width=\linewidth]{TCE_v_esc}
}{
\includegraphics[width=\linewidth]{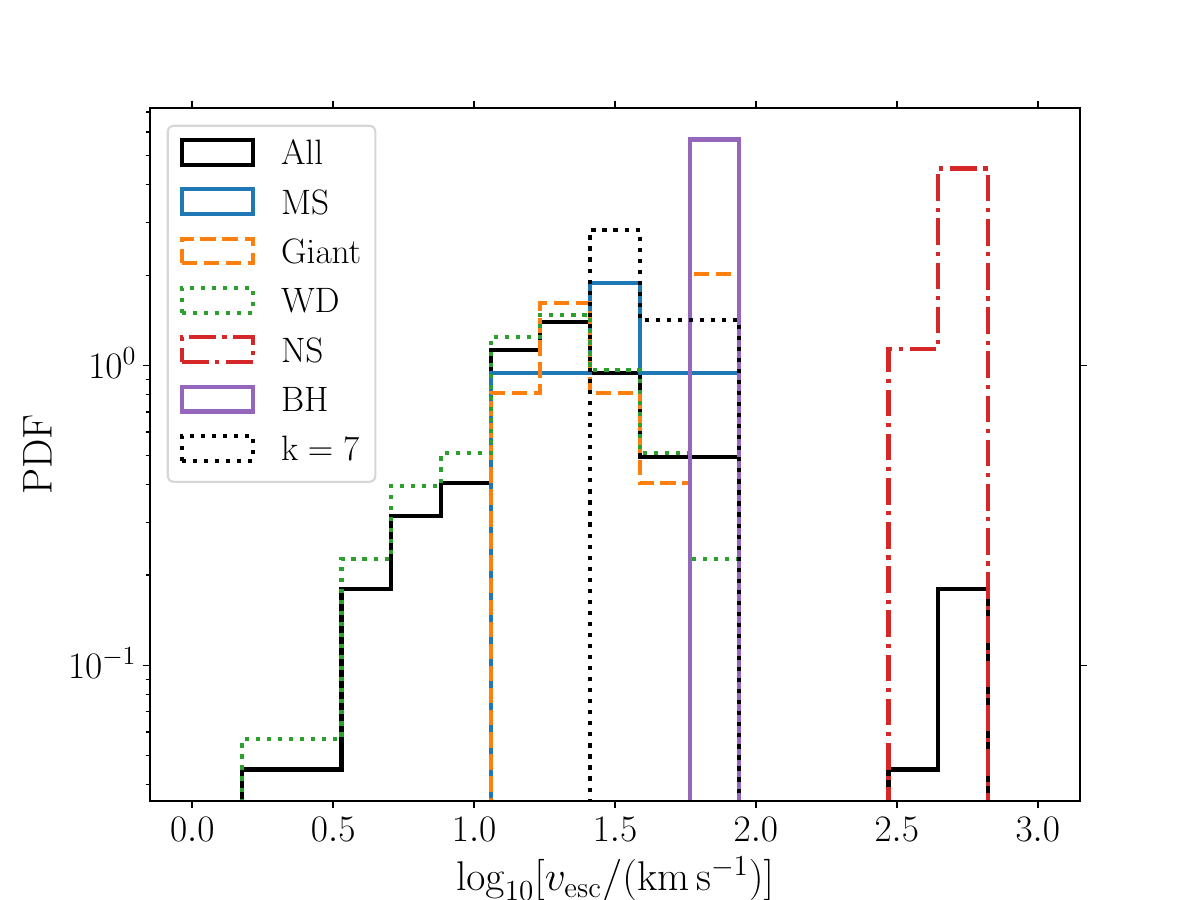}
}
\caption{Distributions of escape speeds (relative to the initial triple centre of mass) for escaping stars following (within $1\,\myr$ of) triple CE evolution. Black solid lines: all unbound objects. Other colors and linestyles indicate distributions for different types of stars. }
\label{fig:v_esc}
\end{figure}

Lastly, we consider the escape speeds of unbound stars following triple CE evolution. \F~\ref{fig:v_esc} shows distributions of escape speeds (relative to the initial triple centre of mass) for escaping stars following (within $1\,\myr$ of) triple CE evolution. In addition to showing the distribution regardless of stellar type (black solid line), we show the distributions for different stellar types (refer to the legend).

The overall escape speed distribution is peaked near $\sim 20\,\kms$. This is significantly larger compared to the escape speed distribution following dynamical instability according to \citet{2021arXiv210713620H}, which peak near $\sim 0.3\,\kms$ and $\sim 1\,\kms$ when a single star, and three stars become unbound, respectively. The on average higher escape speeds in the triple CE case compared to dynamically unstable systems can be understood by noting that the former are typically associated with much more compact triple systems; this implies there exists a larger (orbital) energy reservoir for the escaping star to extract kinetic energy from. The high-velocity tail with $v_\mathrm{esc} \gtrsim 300\,\kms$ is comprised of NSs and BHs for which natal kicks contribute to the escape speed, and the latter depend on our specific assumptions of natal kicks (cf. \S\ref{sect:meth:stel}). A small fraction of WDs attain an escape speed of up to $\sim 100\,\kms$. Although not high enough to be high velocity stars, they can be considered to be intermediate-velocity stars which could be ejected into the Galactic halo \citep[e.g.,][]{2005ApJ...625..838B}.

\begin{figure}
\iftoggle{ApJFigs}{
\includegraphics[width=\linewidth]{TRLOF_stable_delta_m1}
\includegraphics[width=\linewidth]{TRLOF_stable_delta_m2}
\includegraphics[width=\linewidth]{TRLOF_stable_delta_m3}
}{
\includegraphics[width=\linewidth]{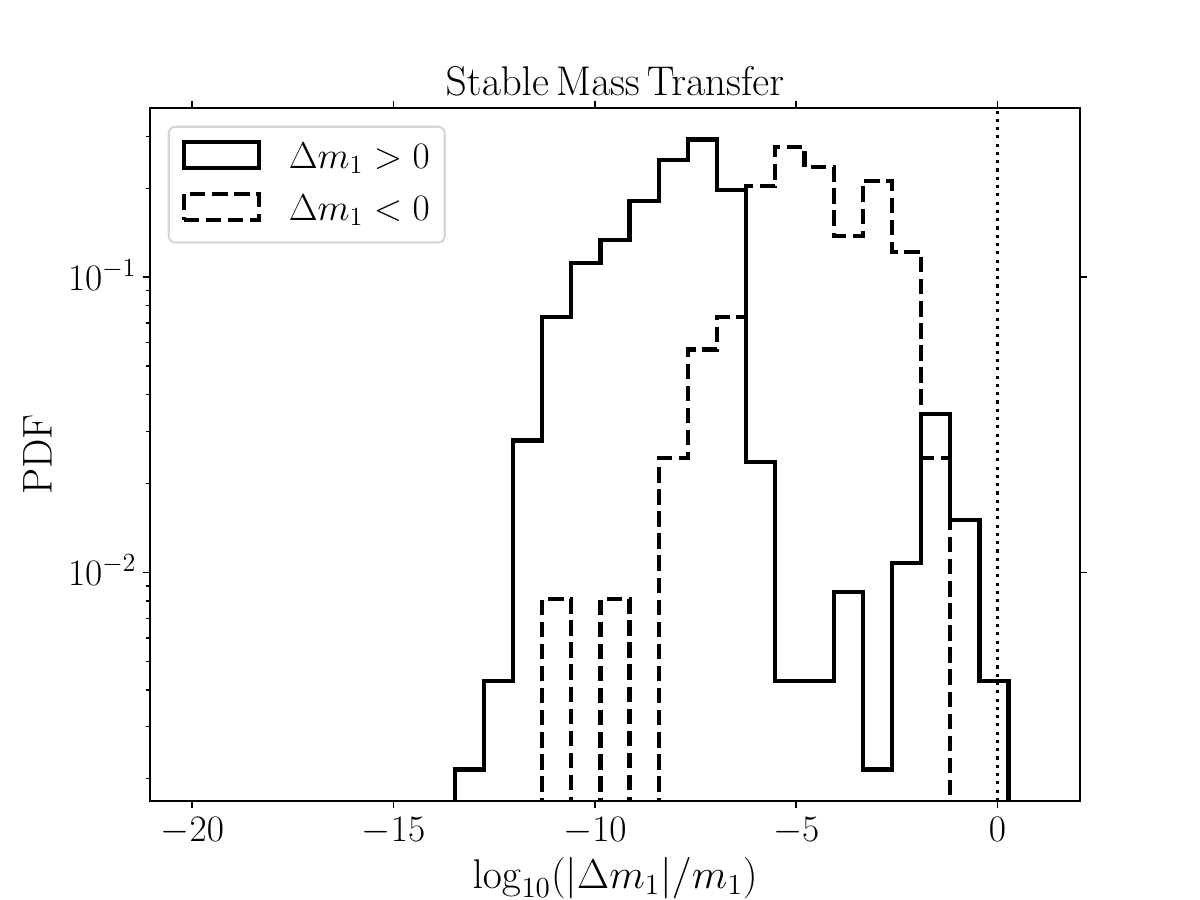}
\includegraphics[width=\linewidth]{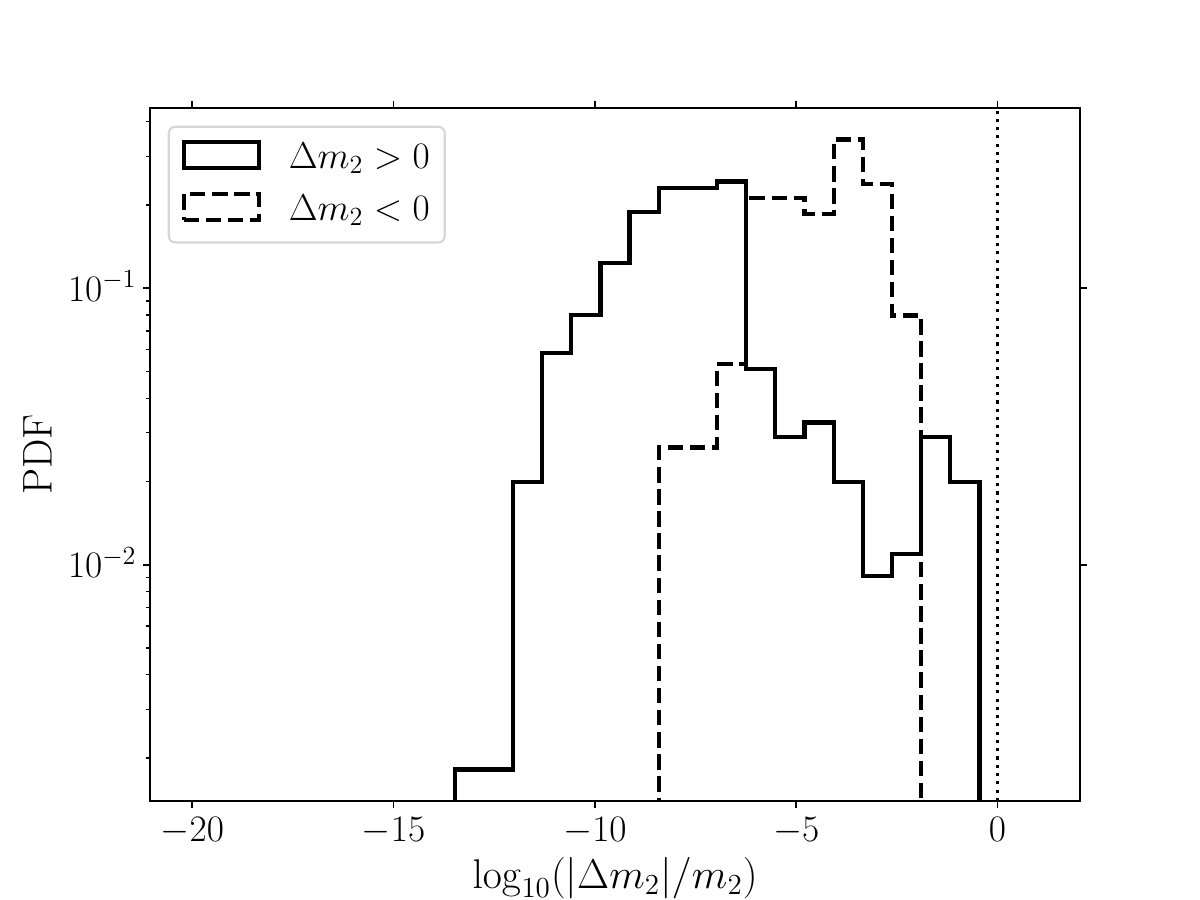}
\includegraphics[width=\linewidth]{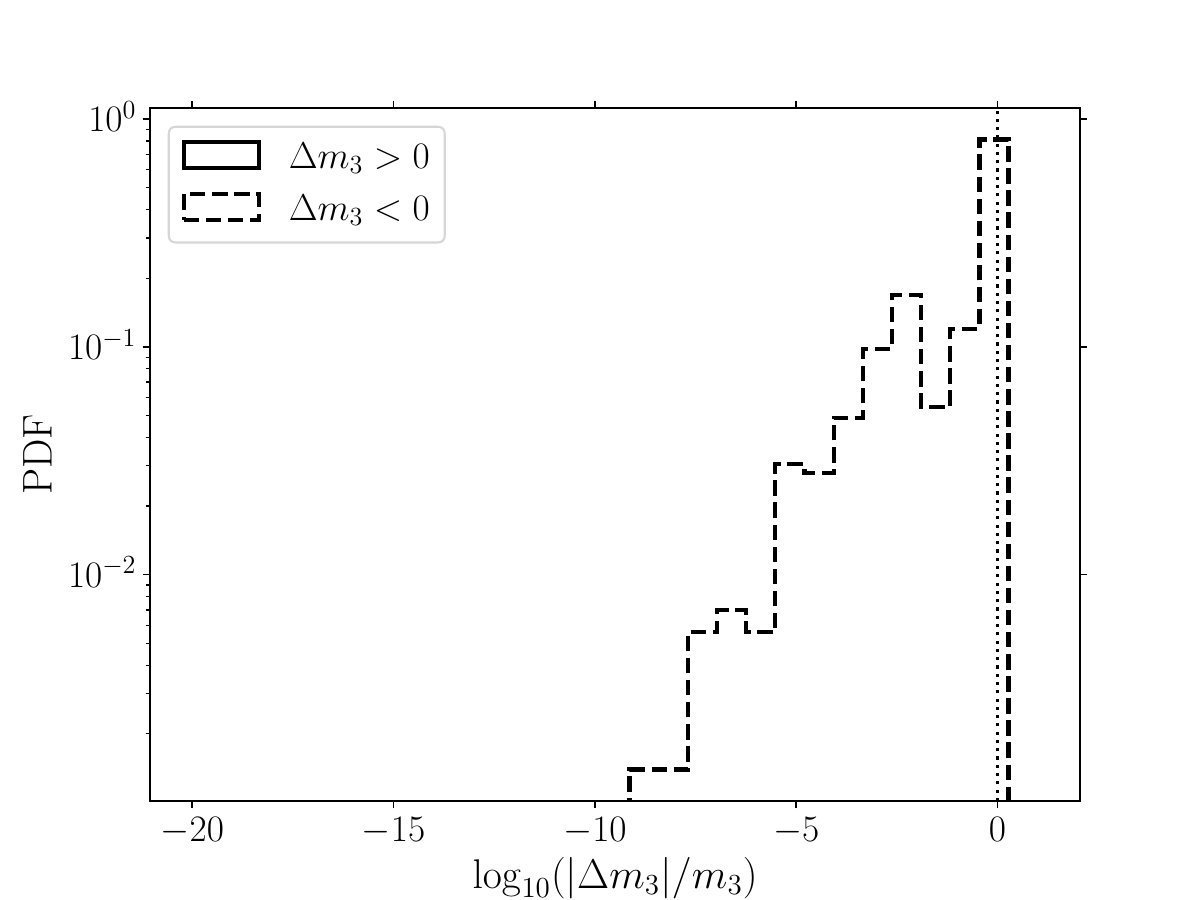}
}
\caption{Fractional changes in the three masses during stable triple RLOF. Distributions for positive (negative) changes are shown with solid (dashed) lines. }
\label{fig:stable_delta_m}
\end{figure}

\subsection{Stable mass transfer}
\label{sect:popsyn:stab}
Here, we focus on properties of systems in which triple RLOF occurs, and mass transfer of the tertiary star onto the inner binary is deemed stable by the \mse~code. \F~\ref{fig:stable_delta_m} shows fractional changes in the three masses during stable triple RLOF. Distributions for positive (negative) changes are shown with solid (dashed) lines.

The inner binary components ($m_1$ and $m_2$) can both increase or decrease during stable triple RLOF, whereas the tertiary star ($m_3$) always decreases in mass. The fractional change in $m_3$ is typically large (peaking near unity), since the tertiary star is not only losing mass due to mass transfer, but it is likely to be evolving quickly and hence significantly losing mass in the form of winds. The decrease in the inner binary components can mainly be attributed to wind mass loss. Mass gain in the inner binary can be due to mass transfer from the tertiary star. Usually, however, the amount of mass accreted is small since the fractional increase is typically $\sim 10^{-8}$ for both components, i.e., the amount of mass accreted is negligible. This can be attributed to the typically short duration of the stable mass transfer phase, peaking at $\sim 0.1 \,\myr$ (see \F~\ref{fig:trlof_times}, red dashed line). The latter can be understood by noting that triple RLOF is usually driven by the expansion of the tertiary star as the latter becomes a giant star. Indeed, the typical tertiary mass for triple RLOF is $m_3\sim 6\,\msun$, and the corresponding Kelvin-Helmholtz timescale is $\sim 1.5\times 10^7\,\yr \, (m_3/\msun)^{-2.5} \simeq 0.17\,\myr$ \citep[e.g.,][]{1994sse..book.....K}. 

An important caveat to the above conclusion is that specific mass accretion efficiencies onto the inner binary were assumed (cf. \S\ref{sect:meth:tr:stab}), and our results are dependent on them. However, given the smallness of the typical transferred amount, even an order-of-magnitude change in the efficiencies would not change our main conclusion that the amount of mass transferred onto the inner binary is often negligible.

\section{Discussion}
\label{sect:dis}

\subsection{General implications}
\label{sect:dis:gen}
As shown by our population synthesis simulations, RLOF of a tertiary star onto the inner binary in triples (i.e., `triple RLOF') is uncommon (occurring in $\sim 0.06\%$ of all triples, cf. Table~\ref{table:fractions}). However, triple RLOF can lead to a wide variety of stellar mergers including MS-MS mergers, mergers involving giant stars, and mergers with compact objects, as shown in Table~\ref{table:rates}, and \F~\ref{fig:trlof_merger_sts}. These mergers could occur within the circumstellar environment of the tertiary donor star, potentially producing peculiar transient events that would otherwise be difficult to explain \citep[e.g.,][]{2021MNRAS.504.5967S}.

\begin{figure}
\iftoggle{ApJFigs}{
\includegraphics[width=\linewidth]{merging_WD_masses_ind}
\includegraphics[width=\linewidth]{merging_WD_masses_tot}
}{
\includegraphics[width=\linewidth]{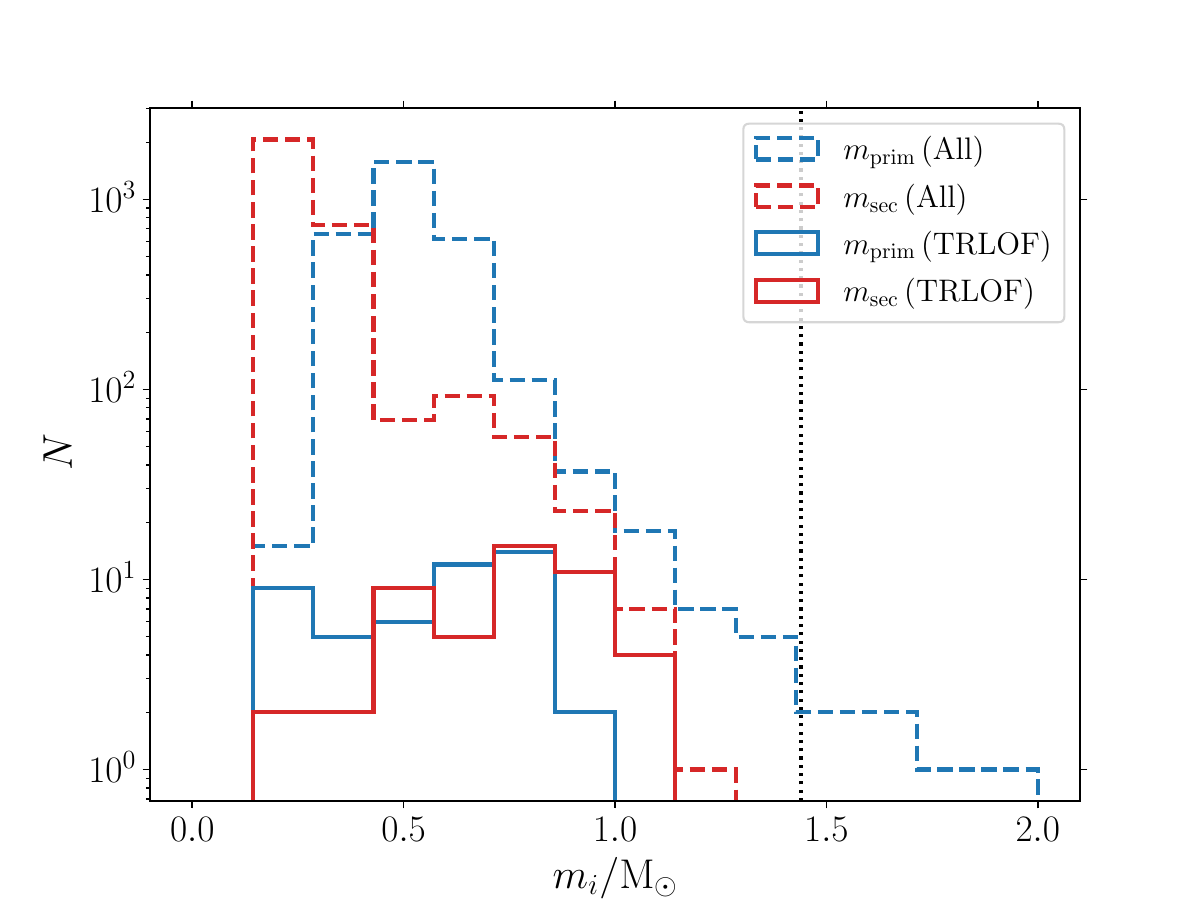}
\includegraphics[width=\linewidth]{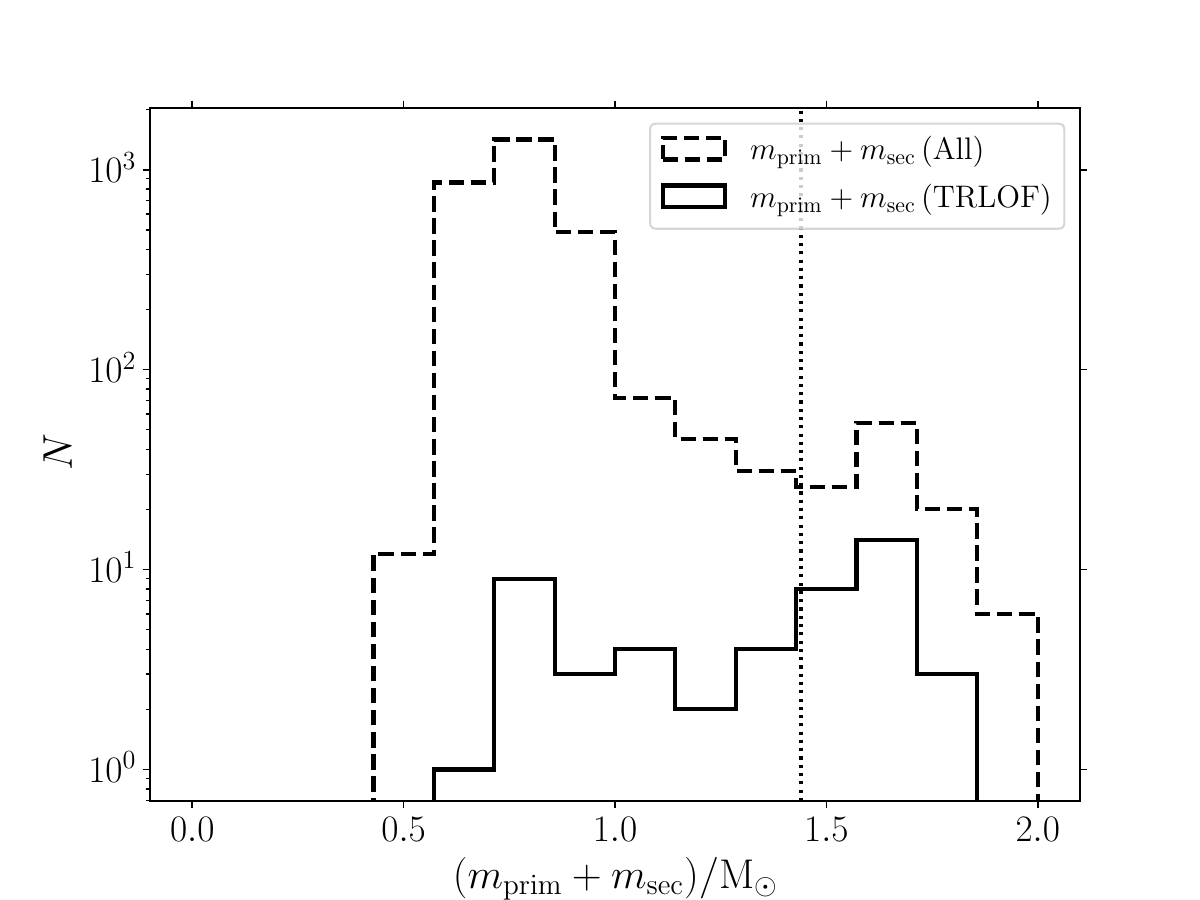}
}
\caption{Distributions of the masses of merging (He and CO) WDs in our simulations. Top panel: primary and secondary masses; bottom panel: total mass. In both panels, the dashed lines correspond to all systems in the population synthesis simulations, whereas the solid lines correspond to those systems involving triple RLOF (both stable and unstable). The black vertical dotted lines indicate the Chandrasekhar mass, $1.44\,\msun$.}
\label{fig:WD_masses}
\end{figure}

\subsection{WD mergers}
\label{sect:dis:WD}
Merging WDs can potentially produce SNe Ia \citep[e.g.,][]{2009ApJ...705L.128R,2010MNRAS.406.2749L,2011A&A...528A.117P,2016ApJ...822...19P}. In particular, physically colliding WDs in extremely eccentric orbits in triples and higher-order systems have been discussed as potentially interesting SNe Ia pathways \citep[e.g.,][]{2011ApJ...741...82T,2012arXiv1211.4584K,2013MNRAS.430.2262H,2013ApJ...778L..37K,2018A&A...610A..22T,2018MNRAS.476.4234F,2018MNRAS.478..620H,2019ApJ...882...24H}. WD collisions, and hence potentially SNe Ia, could also arise after dynamical instability \citep[e.g.,][]{2012ApJ...760...99P,2021arXiv210713620H,2021arXiv210804272T}. 

As shown in the examples (\S\ref{sect:ex}) and in \S\ref{sect:popsyn:frac}, systems undergoing triple RLOF in our simulations can lead to WD mergers. A fraction of $\simeq 4.9 \times 10^{-4}$ of systems in our simulations (this fraction is relative to the `subset' of triples, cf. \S\ref{sect:ICs:mc:sub}) indeed involve (He and CO) WD mergers. This corresponds to a Galactic merger rate of $\simeq 2.6 \times 10^{-6} \, \yr^{-1}$ (cf. Appendix~\ref{app:norm}). Not all of these mergers may lead to SNe Ia, however, since the total mass of the merging WDs does not always exceed the Chandrasekhar mass. Specifically, \F~\ref{fig:WD_masses} shows distributions of the masses of merging (He and CO) WDs in our simulations, with the top panel showing primary and secondary masses, and the bottom panel showing the combined mass of the merging WDs. In both panels, the dashed lines correspond to all systems in the population synthesis simulations, whereas the solid lines correspond to those systems involving triple RLOF (both stable and unstable). 

Interestingly, merging WDs in systems involving triple RLOF typically have somewhat higher combined masses compared to all systems in our simulations, implying that the fraction of systems with a total mass in excess of the Chandrasekhar mass is higher. This can be understood by noting that triple CE systems ($\simeq 71\%$ of WD mergers with triple RLOF led to triple CE evolution) can undergo pre-WD mergers (e.g., a MS-MS merging in the inner binary), raising the remnant WD mass. The pre-WD mass can also be increased by stable triple mass transfer. 

Whether or not a particular WD merger leads to a SNe Ia explosion is uncertain, and comparing the total mass to the Chandrasekhar mass may not give a completely accurate picture in light of the possibility of (sub-luminous) sub-Chandrasekhar-mass SNe \citep[e.g.,][]{2010ApJ...714L..52S,2012ApJ...747L..10P,2013ApJ...770L...8P,2021A&A...649A.155G}. Another important factor to be considered here is the relative velocity and impact parameter, since head-on collisions are expected to more likely produce a SNe Ia explosion \citep[e.g.,][]{2009ApJ...705L.128R,2016ApJ...826..169H}. In any case, even if all of the merging WDs involving triple RLOF in our simulations lead to SNe Ia, we expect the rates to be no higher than $\sim 3 \times 10^{-6}\,\yr$. Although comparable to WD merger rates associated with dynamical stability in triples \citep{2021arXiv210713620H}, the observed Galactic SNe Ia rate is $\sim 10^{-2}\,\yr^{-1}$ \citep{2013ApJ...778..164A}, indicating that triple RLOF interactions likely do not contribute significantly to Galactic SNe Ia. 

\subsection{Caveats and future improvements}
\label{sect:dis:cav}
Mass transfer in stellar systems is generally a complicated process and not even fully understood in binary-star systems, with major uncertainties remaining in, e.g., the stability of mass transfer, and its conservativeness \citep[e.g.,][]{2007A&A...467.1181D}. By extension, mass transfer processes in triple and higher-order systems are even less well understood. Although this work provided some statistical insights into triple RLOF, there are caveats and many areas left unaddressed. 

These caveats include neglecting the extended nature of the inner binary when determining whether or not the tertiary star fills its Roche lobe around the inner binary. Strictly speaking, the Lagrangian points and hence the concept of the Roche lobe no longer apply when the companion is a binary object instead of a point mass (even if all orbits are circular and coplanar). Although the point mass approximation is still useful, a more detailed treatment requires the dynamical Roche lobe to be taken into account \citep{2020MNRAS.491..495D}. 

Furthermore, the boundary between stable and unstable mass transfer when a star fills its Roche lobe around a companion is poorly understood, especially in triples. Here, we adopted a strongly simplified criterion (based on \citealt{2002MNRAS.329..897H}), whereas it is known that such criteria based on population synthesis methods can over-predict the occurrence rate of CE in binaries \citep[e.g.,][]{2011ApJ...739L..48W,2015ApJ...812...40G,2021A&A...650A.107M}. Evidently, these uncertainties are even more severe in the triple case.

Also, our treatment to determine the outcome of triple CE evolution is highly simplified given that the (approximate at best) $\alpha$-CE prescription is extrapolated and applied to the outer orbit, and any subsequent hydrodynamical evolution is ignored. Although the latter may be unimportant in later stages of triple CE evolution \citep{2020MNRAS.498.2957C}, the entire phase can only be accurately modelled with full hydrodynamical simulations \citep{2021MNRAS.500.1921G}. Specifically, \citet{2021MNRAS.500.1921G} found that the inner binary can contribute to the energy budget in triple CE evolution and in particular can accelerate envelope ejection, resulting in a wider orbit compared to the situation when the inner binary in the $\alpha$-CE prescription is treated as a point mass. Therefore, in this work, we may have overestimated the occurrence rate of strong 3-body interactions following triple CE evolution. A further caveat is that, because of current limitations in \mse, a single $N$-body realisation was taken into account when triple CE led to an unstable configuration. However, the orbital phase can be important in determining the outcome of triple CE evolution \citep{2021MNRAS.500.1921G}.

Stable triple mass transfer is also poorly understood, with only few detailed simulations having been carried out (in the stellar context; \citealt{2014MNRAS.438.1909D,2020MNRAS.496.1819L}). Our results relied on simplified prescriptions based on the simulations of \citet{2014MNRAS.438.1909D}. 

Despite the above caveats, we believe that our results, based on currently plausible prescriptions, should be able to provide at least some insights into the qualitative and statistical behavior of triple RLOF evolution. For example, we believe that our result is robust that a triple undergoing triple CE evolution is unlikely to survive the process as a new stable triple system (cf. Table~\ref{table:fractions}).

In future work, some of the above caveats could be addressed by implementing hydrodynamic drag terms within the direct $N$-body simulations, in order to better model the triple CE phase. Although still not a fully self-consistent hydrodynamic treatment (this would currently be too computationally expensive for population synthesis studies), this approach would capture the early phase of triple CE evolution more accurately. Furthermore, the long-term evolution of stable mass transfer evolution could be treated with a test particle approach, in which $N$-body integration is used to determine in a quantitative sense how much mass is expected to be accreted by the inner binary. 

Lastly, we here only considered the `circumstellar' case when the tertiary star fills its Roche lobe around the inner binary. Another possibility is that CE evolution occurs in the inner binary system, and the donor's envelope swells up sufficiently to also engulf the tertiary star \citep[e.g.,][]{2021MNRAS.500.1921G}.

\section{Conclusions}
\label{sect:conclusions}
We have studied statistical aspects of triple Roche lobe overflow (RLOF), i.e., when the tertiary star in a triple-star system fills its Roche lobe around the inner binary. Using the \mse~code \citep{2021MNRAS.502.4479H} which self-consistently models stellar evolution, binary interactions, gravitational dynamics, as well includes simplified prescriptions for triple RLOF interactions, we have quantified which systems in a population of triples are expected to undergo triple RLOF, and studied aspects of both stable triple mass transfer and unstable mass transfer (the latter is assumed to lead to triple CE evolution). We give our main conclusions below. 

\medskip \noindent 1. Based on our simulations, we found that triple RLOF occurs in $\sim 0.06\%$ of all triples. Of these 0.06\%, $\sim 64\%$ of cases to stable mass transfer, and $\sim 36\%$ to triple CE evolution. Triple CE is most often ($\sim 76\%$) followed by one or multiple mergers in short succession, most likely an inner binary merger of two main-sequence stars. Other outcomes of triple CE are a binary+single system ($\sim 23\%$, most of which not involving exchange interactions), and a stable triple ($\sim 1\%$).

\medskip \noindent 2. The overall rate of mergers involving triple RLOF during the evolution of the triple system is $\simeq 1.2 \times 10^{-4}\,\yr^{-1}$, comparable to the rate of mergers in triples following dynamical instability \citep{2012ApJ...760...99P,2021arXiv210713620H,2021arXiv210804272T}. Of these mergers, a significant fraction consists of MS-giant mergers ($\sim 53\%$), MS-MS mergers ($\sim 19\%$), and giant-WD mergers ($\sim 18\%$). 

\medskip \noindent 3. Systems undergoing triple RLOF are typically initially not highly mutually inclined. Initially highly inclined systems are more likely to experience high eccentricities in the inner orbit due to secular evolution and merge before triple RLOF evolution can ensue. Consequently, secular evolution before triple RLOF plays only a small role and the inner orbit eccentricity is typically only slightly enhanced. The outer orbital eccentricity tends to strongly circularise before the onset of triple RLOF, as a result of tidal evolution. 

\medskip \noindent 4. Triple CE (with the tertiary star filling its Roche lobe unstably around the inner binary, i.e., the circumstellar case) can lead to binary+single systems ($\sim 23\%$ of triple CE outcomes). The newly formed binary is typically more eccentric compared to the initial inner binary. Furthermore, the eccentricity distribution of the new binary becomes more consistent with a thermal distribution, as expected for chaotic three-body interactions. 

\medskip \noindent 5. Stable triple RLOF of the tertiary star onto the inner binary typically only leads to a small increase of the inner binary components, if any. The typical fractional increase in the inner binary component masses is $\sim 10^{-8}$. This conclusion is likely not affected by the uncertain accretion efficiency of mass by the inner binary. 

\medskip \noindent 6. Triple CE evolution can produce unbound stars with escape speeds of typically several tens of $\kms$, with up to $\sim 600\,\kms$ for BHs and NSs (the latter result, however, depends strongly on our assumptions on the natal kick velocities). Typical speeds of tens of $\kms$ in the case of triple CE evolution are significantly larger compared to the escape speeds following dynamical instability. In our simulations, a small fraction of WDs attain an escape speed of up to $\sim 100\,\kms$ after triple CE evolution. 
 
\medskip \noindent 7. Mergers of WDs can involve evolutionary pathways with either stable or unstable triple RLOF. However, we estimated the associated Galactic WD merger rate to be $\simeq 2.6 \times 10^{-6} \, \yr^{-1}$. Therefore, even in an optimistic scenario in which all of these mergers lead to SNe Ia, the contribution to the Galactic SNe rate, $\sim 10^{-2}\,\yr^{-1}$ \citep{2013ApJ...778..164A}, is negligible.

\begin{acknowledgements} 
We thank Hagai Perets for comments on the manuscript, and the anonymous referees for helpful reports. A.S.H. thanks the Max Planck Society for support through a Max Planck Research Group.
\end{acknowledgements}

\bibliographystyle{aasjournal}
\bibliography{literature.bib}

\appendix

\section{Normalization and rate calculations}
\label{app:norm}
Here, we briefly describe how events recorded in our population synthesis calculations can be converted into Galactic event rates (this calculation is similar to that of \citealt{2021arXiv210713620H}).

As stated in \S\ref{sect:ICs:mc:sub}, the subset of systems evolved in the population synthesis simulations comprises a fraction $f_\mathrm{calc} \simeq 0.04220$ of all triples underlying our assumed initial distributions (cf. \S\ref{sect:ICs:mc:com}). Next, we estimate the total mass of the complete population of triples. The IMF \eq~(\ref{eq:imf}) implies that a population of $N_\mathrm{s}$ single stars has an average mass of $M_\mathrm{s} \equiv M_\mathrm{Kr} N_\mathrm{s}$, where $M_\mathrm{Kr} \simeq 0.5673 \, \msun$. Assuming a flat mass ratio distribution, a population of $N_\mathrm{bin}$ binary stars has a total mass of $M_\mathrm{bin} \approx (1+\frac{1}{2} ) M_\mathrm{Kr} N_\mathrm{bin} = \frac{3}{2} M_\mathrm{Kr} N_\mathrm{bin}$. Also assuming a flat distribution for the outer mass ratio $q_2 \equiv m_3/(m_1+m_2)$, a population of $N_\mathrm{tr}$ triple stars has a total mass of $M_\mathrm{tr} \approx (\frac{3}{2} + \frac{1}{2} \frac{3}{2} ) M_\mathrm{Kr} N_\mathrm{tr} = \frac{9}{4} M_\mathrm{Kr} N_\mathrm{tr}$. 

Let $\alpha_\mathrm{s}$, $\alpha_\mathrm{bin}$, and $\alpha_\mathrm{tr}$ be the single, binary, and triple fractions, respectively ($\alpha_\mathrm{s} + \alpha_\mathrm{bin} + \alpha_\mathrm{tr} \equiv 1$), and $N_\mathrm{sys}$ the total number of stellar systems (we ignore higher-order multiples such as quadruples in our rate normalization calculation, i.e., we assume that they do not contribute significantly to the total mass). The total mass of a population of single, binary, and triple stars is then approximately
\begin{align}
\nonumber M_\mathrm{tot} & = M_\mathrm{s} + M_\mathrm{bin} + M_\mathrm{tr} \\
\nonumber &\approx \left (N_\mathrm{s} + \frac{3}{2} N_\mathrm{bin} + \frac{9}{4} N_\mathrm{tr} \right ) M_\mathrm{Kr} \\
 &= \left (1 + \frac{1}{2} \alpha_\mathrm{bin} + \frac{5}{4} \alpha_\mathrm{tr} \right) N_\mathrm{sys} M_\mathrm{Kr}.
\end{align}
We assume mass-independent fractions $\alpha_\mathrm{bin} = 0.6$, and $\alpha_\mathrm{tr} = 0.1$. 

The number of systems calculated in our simulations is given by $N_\mathrm{calc} = f_\mathrm{calc} \alpha_\mathrm{tr} N_\mathrm{sys}$. For a particular outcome $X$ with $N_X$ systems in the simulations, the corresponding Galactic rate $R_X$ (with a Galactic star formation rate of $R_\mathrm{SFR}$) is then given by
\begin{align}
\nonumber R_X &= N_X \frac{R_\mathrm{SFR}}{M_\mathrm{tot}} = \frac{N_X f_\mathrm{calc} \alpha_\mathrm{tr} R_\mathrm{SFR}}{N_\mathrm{calc} \left (1 + \frac{1}{2} \alpha_\mathrm{bin} + \frac{5}{4} \alpha_\mathrm{tr} \right) M_\mathrm{Kr}} \\
&\simeq 0.005220 \, \yr^{-1} \, \left ( \frac{N_X}{N_\mathrm{calc}} \right )\left ( \frac{R_\mathrm{SFR}}{1 \, \msun \, \yr^{-1}} \right ),
\end{align}
where we substituted for the numerical value a Galactic star formation rate of $R_\mathrm{SFR} = 1\,\msun\,\yr^{-1}$ \citep{2010ApJ...710L..11R}. We remark that $N_X/N_\mathrm{calc} = f_X$ is simply the fraction of systems in the population synthesis calculations corresponding to outcome $X$.

\end{document}